\documentclass[paper,preprint,12pt]{JHEP}
\usepackage{epsfig}
\def\ben{\begin{enumerate}}  \def\een{\end{enumerate}}
\def\bit{\begin{itemize}}    \def\eit{\end{itemize}}
\def\beq{\begin{equation}}   \def\eeq{\end{equation}}
\def\beeq{\begin{eqnarray}}  \def\eeeq{\end{eqnarray}}
\def\bq{\begin{quote}}       \def\eq{\end{quote}}
\def\a0{\bar\alpha_0}
\def\ae{\alpha_{\mbox{\scriptsize eff}}}

\def\as{\alpha_{\mbox{\tiny S}}}
\def\b0{\beta_0}
\def\cN{{\cal N}}

\def\Ecm{E_{\mbox{\scriptsize cm}}}
\def\ee{e^+e^-}

\def\jet{{\mbox{\scriptsize jet}}}
\def\kij{k^2_{\bot ij}}
\def\kp{k_\perp}

\def\kt{k_\bot}
\def\lms{\Lambda^{(5)}_{\overline{\mbox{\tiny MS}}}}

\def\mx{{\mbox{\scriptsize max}}}
\def\NP{{\mbox{\tiny NP}}}

\def\st{\sigma_{\mbox{\scriptsize tot}}}
\def\ycut{y_{\mbox{\tiny cut}}}
\def\frac#1#2{ {{#1} \over {#2} }}

\def\VEV#1{\left\langle #1\right\rangle}
\def\cO#1{{\cal{O}}\left(#1\right)}
\def\beq{\begin{equation}}
\def\beeq{\begin{eqnarray}}
\def\eeq{\end{equation}}
\def\eeeq{\end{eqnarray}}
\def\cpc#1#2#3{{\it Comp.\ Phys.\ Commun.\ }{\bf #1} (19#3) #2}
\def\jpg#1#2#3{{\it J.\ Phys.\ }{\bf G#1} (19#3) #2}
\def\np#1#2#3{{\it Nucl.\ Phys.\ }{\bf B#1} (19#3) #2}
\def\pl#1#2#3{{\it Phys.\ Lett.\ }{\bf B#1} (19#3) #2}
\def\pr#1#2#3{{\it Phys.\ Rev.\ }{\bf D#1} (19#3) #2}
\def\prl#1#2#3{{\it Phys.\ Rev.\ Lett.\ }{\bf #1} (19#3) #2}
\def\zp#1#2#3{{\it Z.\ Physik }{\bf C#1} (19#3) #2}
\def\Ord{\buildrel{\scriptscriptstyle <}\over{\scriptscriptstyle\sim}}

\title{Better Jet Clustering Algorithms}
\author{Yu.L.\ Dokshitzer\thanks{Permanent address:
St Petersburg Nuclear Physics Institute, Gatchina, St Petersburg 188350,
Russian Federation.} \\
INFN Sezione di Milano, Via Celoria 16, 20133 Milan, Italy\\
E-mail: \email{yuri@mi.infn.it}}
\author{G.D.\ Leder, S.\ Moretti, B.R.\ Webber \\
        Cavendish Laboratory, University of Cambridge,\\
        Madingley Road, Cambridge CB3 0HE, U.K.\\
        E-mail: \email{leder@hep.phy.cam.ac.uk}, etc.}

\abstract{We investigate modifications to the $\kt$-clustering
jet algorithm which preserve the advantages of the original Durham
algorithm while reducing non-perturbative corrections and providing
better resolution of jet substructure.  We find that a simple change in
the sequence of clustering (combining smaller-angle pairs first),
together with the `freezing' of soft resolved jets, has
beneficial effects.}
\preprint{Cavendish--HEP--97/06}
\begin{document}
\section{Introduction}
\label{sec:intro}
Jet clustering algorithms have become an indispensable tool for the
analysis of hadronic final states in $\ee$ annihilation. More recently
they have also started to be applied to other types of particle collisions.
Clustering algorithms have permitted a wide range of important tests of QCD
and will be of continuing value in more refined studies and in searches for
new physics. It therefore remains worthwhile to think of ways in which the
existing algorithms can be modified to improve their theoretical
properties and phenomenological performance.

In the present paper we propose improvements which can be applied to
any iterative clustering algorithm of the basic JADE type \cite{JADE},
although we mainly discuss them with reference to the so-called Durham
or $\kt$ variant of that algorithm \cite{Durham}. By the JADE
type of algorithm we mean an exclusive iterative pairwise clustering
scheme, in which jets are constructed out of primary objects,
the latter being hadrons or calorimeter cells in the real experimental
case and partons in the perturbative theoretical calculation.  The term
{\em exclusive} means that each primary object is assigned to a
unique jet and each final state has a unique jet multiplicity, for a
given value of the jet resolution parameter $\ycut$.

Algorithms of the JADE type have two basic components:

1) a test variable $y_{ij}$, and

2) a combination procedure.

The test variable is used to decide whether the objects $i$ and $j$ should
be combined, according to whether $y_{ij}<\ycut$. It is also used to
choose which objects to consider next for combination, namely the pair
with the smallest value of $y_{ij}$. In the original JADE algorithm,
$y_{ij}=M^2_{ij}/Q^2$ where $Q$ is the hard scale (i.e.\ the centre-of-mass
or visible energy in $\ee$ annihilation) and
\beq\label{Mijdef}
M^2_{ij} = 2E_i E_j(1-\cos\theta_{ij})\;,
\eeq
which is essentially the invariant mass-squared of the pair. The combination
procedure specifies the properties of the new object formed by
combining $i$ and $j$, for example that its four-momentum should
be simply the sum $p_{ij}=p_i+p_j$ (the so-called E scheme,
which we shall adopt in the present paper). The
clustering procedure is repeated until no objects can be combined further
(all $y_{ij}>\ycut$), at which stage all objects are defined as jets.

A crucial point is that the {\em identical} algorithm should be applicable
to real experimental data and to the partons that appear in the
perturbative calculation. It follows that an essential feature of the
algorithm must be {\em infrared safety}, i.e.\ insensitivity to the
emission of arbitrarily soft and/or collinear particles. Otherwise,
the presence of massless partons in the perturbative calculation would
lead to divergent results. Beyond this fundamental feature, the ideal
algorithm should lead to a close correspondence between the theoretical
(partonic) and actual (hadronic) jet characteristics and multiplicities
after clustering, over the widest
possible range of values of $\ycut$. That is, it should have the
smallest possible {\em hadronization corrections}, this being the
name given to empirical adjustments that are applied to the
theoretical predictions before comparing them with experiment.
One would also like the algorithm to have good theoretical
properties, in particular resummability of large terms to all
orders in perturbation theory and a reduced renormalization-scale
dependence in fixed order.

While the original JADE algorithm satisfies the basic criteria of infrared
safety, its theoretical properties turn out to be surprisingly complicated.
This is because the invariant mass is not the most relevant variable
for the evolution of QCD jets, and consequently the multijet phase
space has a non-factorizing structure when expressed in term of
the JADE test variable \cite{BS1,Stef}. As a result the range of
theoretical predictions that can be compared with experiment
using this algorithm is limited, and the hadronization corrections
are not optimally small.

A basic weakness of the JADE algorithm is illustrated by the way it
deals with the parton-level `seagull diagram' (Fig.~\ref{fig_seagull}),
in the phase-space region where the two gluons $g_3$ and $g_4$ are
soft and almost collinear with the quark and antiquark \cite{BS1}.
\FIGURE[htb]{
\epsfig{file=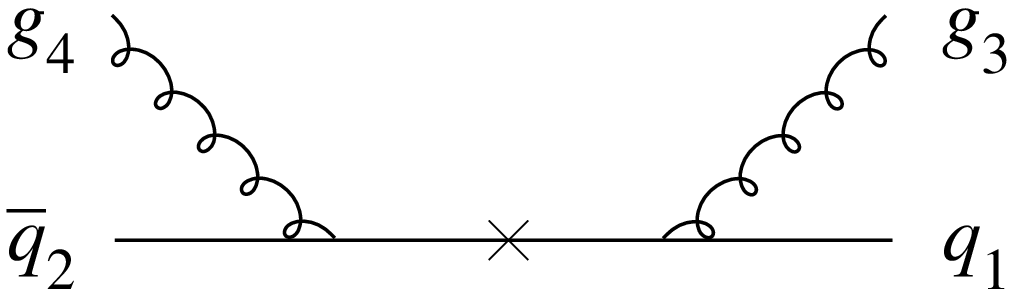,height=3cm}
\caption{The `seagull diagram'.}
\label{fig_seagull}
}
We then have
\beq\label{Jseagull}
y_{13}\sim x_3\theta_{13}^2\;,\;\;\;\; 
y_{24}\sim x_4\theta_{24}^2\;,\;\;\;\; 
y_{34}\sim x_3 x_4\;,
\eeq
where $x_i=2E_i/Q$.  Since all these are of the same order in small
quantities, there is an important subregion in which $y_{34}$ is
the smallest. In that subregion the two gluons will be combined
first, making a `phantom' gluon jet with a resultant momentum
at a large angle to the quark and antiquark, where there are
in fact no particles.

The shortcomings of the JADE algorithm are greatly alleviated by the
Durham modification \cite{Durham,CDOTW,BS2}, which consists
of replacing $M^2_{ij}$ in the definition of the test variable by
$\kij$, where
\beq\label{ktijdef}
\kij = 2\min\{E_i,E_j\}^2(1-\cos\theta_{ij})\;,
\eeq
which is essentially the relative transverse momentum-squared of $i$
and $j$.\footnote{Historically, the first algorithm based on a similar
variable ({\tt LUCLUS}) was developed and included in the JETSET
program by Sj\"ostrand \cite{Jetset}.}
This choice of test variable reflects the more fundamental role
of the transverse momentum in setting the scale of jet evolution, as
the argument of the running coupling, and in defining the boundary
between perturbative and non-perturbative physics. In the treatment
of the seagull diagram, Eqs.~(\ref{Jseagull}) become
\beq\label{Dseagull}
y_{13}\sim x_3^2\theta_{13}^2\;,\;\;\;\; 
y_{24}\sim x_4^2\theta_{24}^2\;,\;\;\;\; 
y_{34}\sim \min\{x_3^2, x_4^2\}\;,
\eeq
and so the two gluons cannot be combined first in this
region of phase space.  As a result, one finds that
the phase space now has simple
factorization properties and the available range of predictions
is increased. In particular, the terms in the perturbation series
that involve leading and next-to-leading powers of $\ln\ycut$ can be
identified and resummed to all orders \cite{CDOTW,BS2}, thereby improving
the reliability of the predictions at small values of $\ycut$.
Furthermore, hadronization corrections, estimated according to the
best available Monte Carlo models \cite{Jetset, Herwig}, are found
to be reduced \cite{Opalclus,BKSS}.

The good features of the Durham algorithm at low $\ycut$
lead one to hope that it should be possible to probe
the interface between perturbative and non-perturbative QCD
by studying jet properties as a function of $\ycut$ in the
relevant region, say $\ycut <10^{-4}$ at $Q=M_Z$,
corresponding to $\kt <1$ GeV.  One would like, for
example, to see whether the low-energy behaviour
of the effective strong coupling could be studied
in this way \cite{DKT96,DLMWprep}.
However, that hope is dashed by the observation \cite{CDFW1,OPALnj}
that the hadronization corrections, although reduced,
are still substantial at $\ycut\sim 10^{-3}$,
corresponding to $\kt \sim 3$ GeV, well above the
values at which non-perturbative effects would normally
be expected.

In the following Section we discuss the defects of the Durham
algorithm which lead to this situation, and the steps that can be
taken to remedy them. We define modified algorithms and study the
associated non-perturbative effects that would be expected in
simple hadronization models. In Sect.~\ref{sec:fixedorder}
we compute the jet fractions
for these new algorithms to next-to-leading order in perturbation
theory and study their properties. Sect.~\ref{sec:njets} is devoted
to studies of the mean jet multiplicity, first in resummed perturbation
theory and then using the HERWIG event generator \cite{Herwig}.
Comparisons show clearly the benefits of the new algorithms
at small values of $\ycut$, beyond the reach of fixed-order perturbation
theory. Finally, in Sect.~\ref{sec:conc}
we briefly summarize our results and conclusions.

\section{Jet algorithms: defects and cures}
\label{sec:algo}
\subsection{`Junk-jet' formation}
The basic problem that we have to address is as follows.
With decreasing $\ycut$, all jet algorithms inevitably start
to search for jets amongst the hadrons with low transverse momenta
and to form spurious `junk-jets' from them,
pretending that these are legitimate, resolvable gluon jets.
As we shall see below, the original JADE algorithm, with
its test variable related to invariant mass, accumulates the first
junk-jet at $\ycut\sim\lambda/Q$ , where $\lambda\sim 0.5$ GeV is a soft
scale set by the mean non-perturbative transverse momentum. The Durham
algorithm eventually falls into the same trap, although at
lower $\ycut$.

One might expect a $\kp$-based algorithm to avoid resolving junk-jets
as long as $\ycut$ is kept above $\lambda^2/Q^2$. The Durham algorithm,
however, does not respect such a natural expectation. According to the
clustering procedure adopted, one usually starts from the softest particle
in a jet (call it hadron \#1) and merges it with the one nearest in
{\em angle}, to minimize the relative $\kp$. Thus hadron \#1
gets clustered not with the leading hadron in the jet but, typically,
with the softest among the hadrons which happen to lie on the same side
of the jet axis in the transverse plane. Half of them do, on average.
The typical transverse momentum of the largest junk-jet, and hence
the $\ycut$ value at which it is resolved, is therefore enhanced by
a factor proportional to $N(Q)$, the soft hadron multiplicity 
in a jet at scale $Q$, which is proportional to $\ln^p Q$, with $p$ depending
on the model of soft physics. As a result, junk-jets start to appear
at $\ycut \propto \ln^{2p} Q (\lambda/Q)^2$. Taking account of fluctuations,
such a misinterpretation proves to be a substantial effect.

To understand these features in more detail, we can examine
the action of the algorithms on simple models of the
non-perturbative hadronization process. Consider for example
the simplest possible string or `tube' model, in
which the hadronization of a back-to-back two-parton system
yields a distribution of hadrons which is uniform in rapidity
$\eta$ and strongly damped in transverse momentum $p_t$, both
variables being defined with respect to the original parton
directions. If the particle density in $(\eta,p_{tx},p_{ty})$-space
is given by $\rho(p_t)$, we have (neglecting hadron masses)
\beq\label{Qtube}
Q=\Ecm = \int d\eta d^2 p_t \rho(p_t) p_t \cosh\eta = 2\lambda\sinh Y
\eeq
where $Y\sim \ln(Q/\lambda)$ is the maximum value of $|\eta|$, and
\beq\label{mudef}
\lambda \equiv \int d^2 p_t \rho(p_t) p_t = N\VEV{p_t}/2Y
\eeq
where the total hadron multiplicity in this model is
$N\sim (2\lambda/\VEV{p_t})\ln(Q/\lambda)$.

Ideally, we would like hadronization to affect the jet structure of events
as little as possible, and therefore we would prefer the tube to remain
classified as a two-jet final state down to the smallest possible
values of the resolution $\ycut$.  Nevertheless any
algorithm will eventually resolve a third (junk-) jet at some
value $\ycut=y_3$; the algorithm should be designed to make this
as small as possible.

The action of various algorithms in resolving a third jet inside the
tube is illustrated in Fig.~\ref{fig_tubes}. In the JADE
algorithm, the resolution measure is related to the
invariant mass, and therefore jets tend to be resolved as slices of
the tube. In particular, an extra third jet is resolved when
$\ycut\sim M^2_\jet/Q^2$, where $M_\jet$ is the mass of the jet
in either hemisphere, given by
\beq\label{Mtube}
M^2_\jet = E^2_\jet - p^2_\jet =2\lambda^2\cosh Y\sim \lambda Q\;,
\eeq
so that
\beq\label{yJtube}
\VEV{y_3}^J\sim\frac \lambda Q\;.
\eeq
\FIGURE[t]{
\epsfig{file=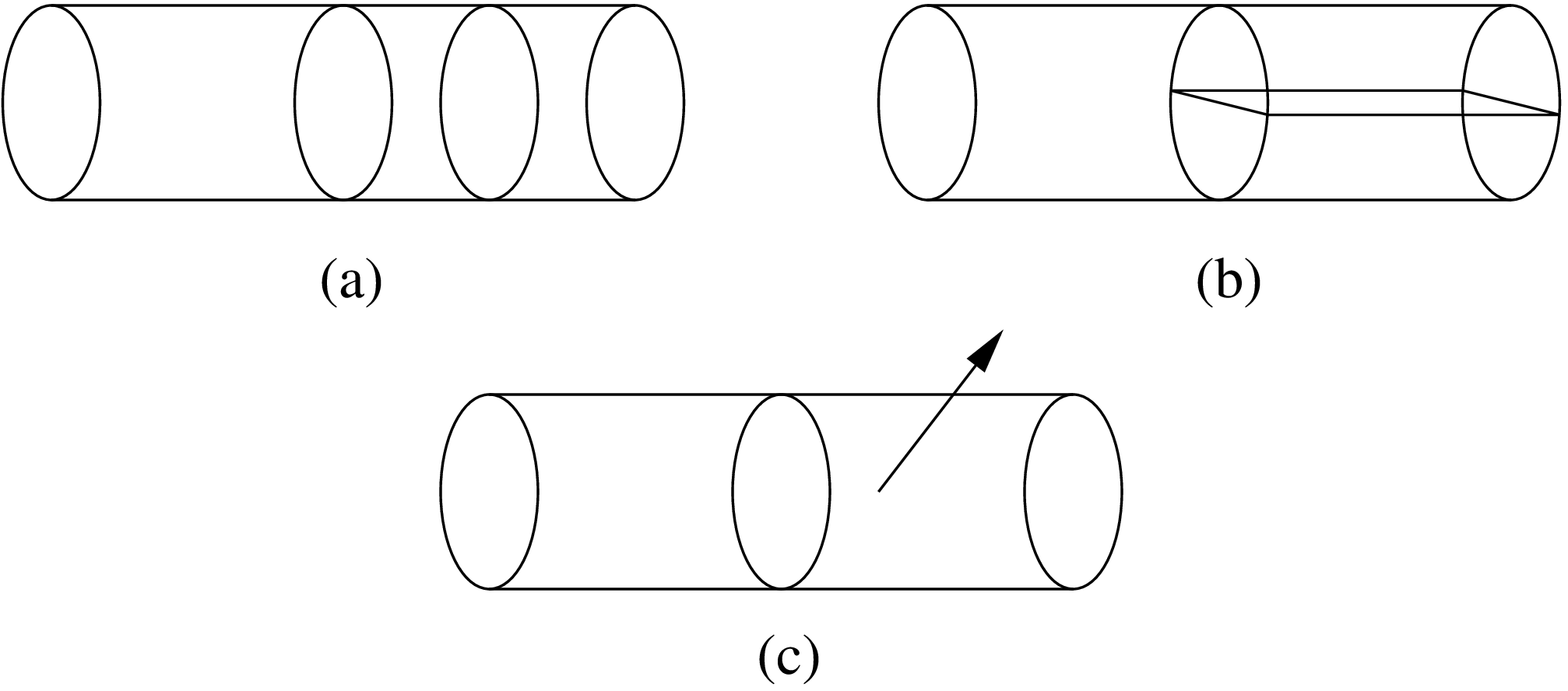,height=16cm,angle=270}
\caption{Resolving a third jet in the final state of the tube model:
(a) JADE, (b) Durham, (c) angular-ordered Durham algorithm.}
\label{fig_tubes}
}
This is shown by the dashed curve in Fig.~\ref{fig_tube3}, to be compared
with data from a Monte Carlo simulation of the tube model (squares).
For this simulation, we simply generated a number $N$ of massless
four-momenta with an exponential transverse momentum distribution and
a uniform rapidity distribution in the interval $-Y<\eta<Y$,
$Y$ being given by Eq.~(\ref{Qtube}) and $N$ by Eq.~(\ref{mudef}).
As illustrative values, we have taken $\lambda=0.5$ GeV
and $\VEV{p_t}=0.3$ GeV.

\FIGURE[t]{
\epsfig{file=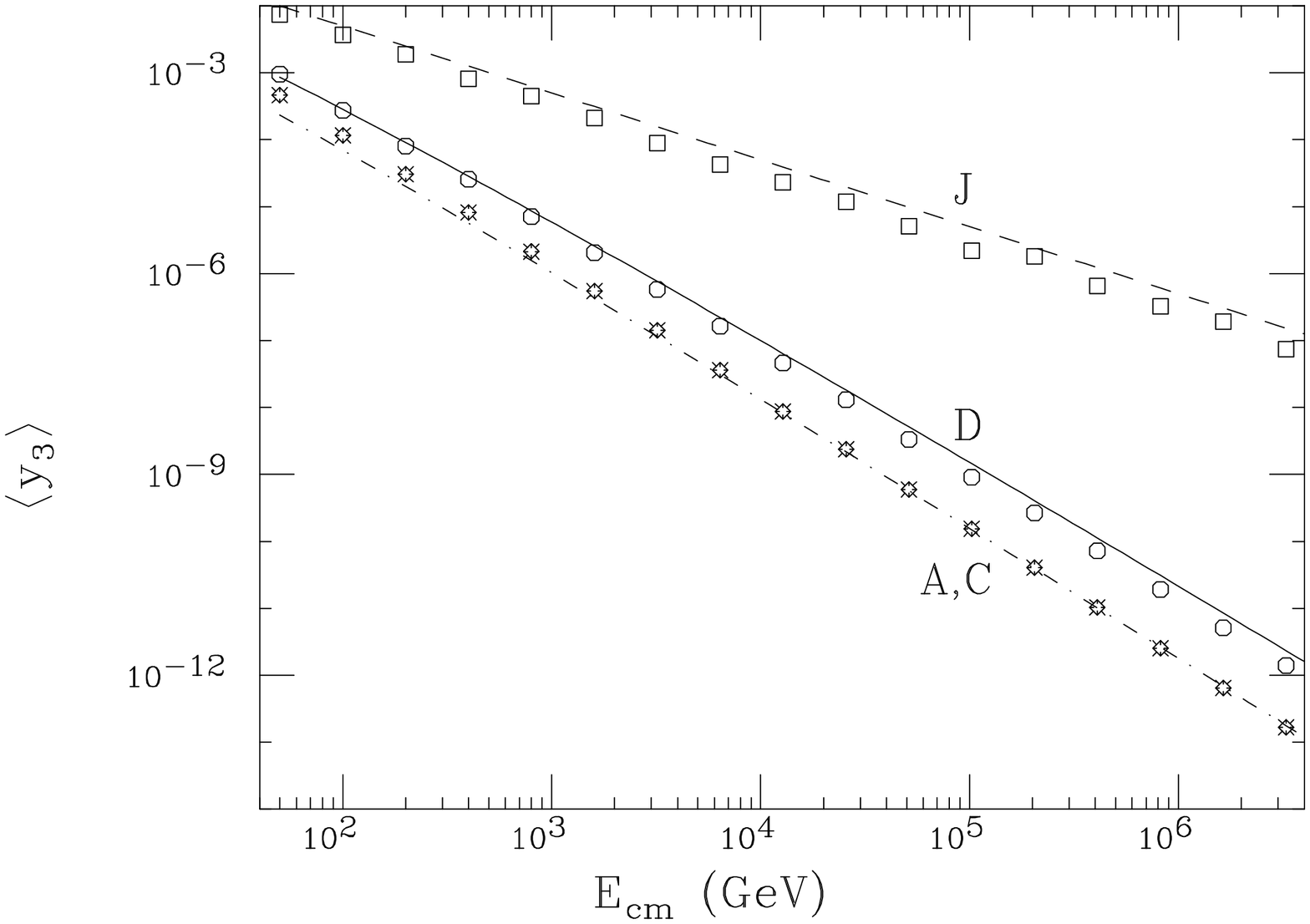,height=16cm,angle=90}
\caption{Mean values of the three-jet resolution in the `tube'
hadronization model, for the J (square symbols), D (circles) and
A, C (stars) jet clustering schemes. The corresponding curves
show the approximate formulae discussed in the text.}
\label{fig_tube3}
}

In the original Durham (D) algorithm, to resolve a third jet one has
to subdivide one of the jets into two parts with relative transverse
momentum $P_t$, where
\beq\label{Pttube}
P_t^2\sim \ycut Q^2\;.
\eeq
The largest value of $\ycut$ at which this can be achieved occurs
when one half of the tube is divided axially into two half-cylinders,
as illustrated in Fig.~\ref{fig_tubes}(b), giving
\beq\label{Ptjet}
P_t\sim \int_0^Y\int d^2 p_t |p_{tx}| \rho(p_t)\sim \frac 2 \pi \lambda Y
\eeq
and hence
\beq\label{yDtube}
\VEV{y_3}^D\sim\left(\frac{2\lambda\ln(Q/\lambda)}{\pi Q}\right)^2\;.
\eeq
This is shown by the solid curve in Fig.~\ref{fig_tube3}, which
agrees well enough with the Monte Carlo tube model data (circles).
We see the expected great improvement relative to the JADE algorithm,
due to the power-suppression factor of $1/Q^2$ rather than $1/Q$.
However, the presence of the log-squared enhancement factor
means that the coefficient of $1/Q^2$ is far larger than
$\cO{\lambda^2}$, the order of magnitude that one 
might hope to be achievable with an optimal jet algorithm.

An alternative way of estimating non-perturbative contributions
to $\VEV{y_3}$ has been proposed in Ref.~\cite{BPY}.
At lowest order in perturbation theory, for any infrared-safe
jet algorithm, this quantity is proportional to $\as$.
In higher orders it is given by a power series in $\as(Q)$,
where the argument of the coupling is set by the only
available hard-scattering scale $Q=\Ecm$.  Now although the
perturbative predictions may be expressed in terms of $\as(Q)$,
one cannot avoid sensitivity to the region of low
momenta $k\ll Q$ inside
integrals that contribute to those predictions.
This sensitivity makes the perturbation series in
$\as(Q)$ strongly divergent at high orders, leading to
power-behaved ambiguities.

In the `dispersive approach' of Ref.~\cite{BPY} these
so-called renormalon ambiguities are resolved by
assuming the existence of a universal low-energy
effective strong coupling $\ae(k)$ with sensible
analyticity properties. This leads to genuine
power-behaved non-perturbative contributions,
which may be parametrized in terms of moments of
$\delta\ae(k)$, the discrepancy between the effective
coupling and its perturbative expansion in terms of $\as(Q)$
(up to the order included in the perturbative contribution).

As was observed in Ref.~\cite{BPY}, the expectation based on
the dispersive approach is that $\VEV{y_3}$ should have a
leading non-perturbative contribution proportional to $1/Q$
in the case of the JADE algorithm and proportional to $\ln Q/Q^2$
for the Durham algorithm. Thus the predicted
power behaviour agrees in each case with that expected
from the simple tube model, although the logarithmic
enhancement of the non-perturbative contribution for
the Durham algorithm is not as large as in the tube model.

In Sect.~\ref{sec:njets} we shall present the results of
studies of hadronization effects in a more realistic model.

\subsection{Angular-ordered Durham (A) algorithm}
\label{subsec:Aalg}
We now show that a simple modification of the Durham algorithm
suffices to delay the onset of junk-jet formation, which results
mainly from a non-optimal sequence of clustering, rather than from
a poor definition of the test variable (as was the case for the
JADE algorithm). The key to alleviating the problem is
to notice that the most general definition of a clustering
algorithm in fact involves {\em three components}
rather than the two mentioned above, viz.  

0) an ordering variable $v_{ij}$,

1) a test variable $y_{ij}$, and

2) a combination procedure.
  
In other words, we now distinguish between the variable $v_{ij}$,
used to decide which pair of objects to test first, and
the variable $y_{ij}$ to be compared with the resolution parameter
$\ycut$. The algorithm then operates as follows. One considers first
the pair of objects $(ij)$ with the smallest value of $v_{ij}$.
If $y_{ij}<\ycut$, they are combined. Otherwise the pair with the
next smallest value of $v_{ij}$ is considered, and so on until
either a $y_{ij}<\ycut$ is found or, if not, clustering has finished
and all remaining objects are defined as jets.

From this viewpoint we see that JADE-type ($v_{ij}\equiv y_{ij}$)
algorithms, including the Durham variant, are likely to produce the
{\em largest} number of jets for a
given definition of the test variable and combination procedure.
Suppose that at some stage of clustering the pair $(ij)$ with the
smallest value of $y_{ij}$ does not have the smallest value of $v_{ij}$.
We can assume that $y_{ij}<\ycut$, since otherwise clustering has
already finished. If the pair $(kl)$ with the smallest value of $v_{kl}$
has $y_{kl}<\ycut$, they will be combined. Otherwise, the pair with the
next smallest value will be considered, and so on until $v_{ij}$ is
reached, whereupon $i$ and $j$ will be combined.  Thus at least the
$(ij)$ clustering performed by the JADE-type algorithm is available
at any stage. However, we cannot make a rigorous argument that the
JADE-type jet multiplicity is an upper bound on that for
$v_{ij}\neq y_{ij}$ algorithms, because the sequence of
clustering will be different in general.

Once one separates the functions of the ordering and test variables,
it is natural to adopt the {\em relative angle} for ordering purposes,
thus taking into account the angular-ordering property of QCD \cite{AO}.
Recall that angular ordering corresponds to the fact that a soft gluon
emitted at some angle to the jet axis cannot resolve the colours of
jet constituents at smaller angles. For a correct correspondence with
the colour structure of the jet, one should therefore combine the
constituents at smaller angles first. We shall refer to this choice
of ordering variable, combined with the Durham test variable and
combination procedure, as the {\em angular-ordered Durham}
(A) algorithm. One starts with a table of the energies $E_i$
of primary objects and their relative angles
$\theta_{ij}$, or equivalently the quantities
\beq\label{vijdef}
v_{ij} = 2(1-\cos\theta_{ij})\;,
\eeq
all defined in the $\ee$ c.m.\ frame.
The A algorithm is then as follows:

\vspace{1ex}
\noindent{\bf Step 1}. Select the pair of objects ($ij$) with
the minimal value of the ordering variable, $v_{ij}$.

\noindent{\bf Step 2}. Inspect the value of the test variable,
$$y_{ij} = \min\{E_i,E_j\}^2\,v_{ij}\;.$$
\bit
\item If $y_{ij}< \ycut$, then update the table by deleting $i$ and $j$,
introducing a new particle ($ij$) with 4-momentum $p_{ij}=p_i+p_j$, and
recomputing the relevant values of the ordering variable;
then go to Step 1. 
\item  If $y_{ij}\ge \ycut$, then consider the pair (if any) with the next
smallest value of the ordering variable and repeat Step 2. If no such
pair exists, then clustering is finished and all remaining objects are jets.
\eit

\vspace{1ex}
The earlier discussion of the logarithmic enhancement of non-perturbative
effects in the Durham algorithm showed that it arises because
spurious junk-jets can be formed by combining soft, large-angle
(low-rapidity) hadronization products. The A algorithm
avoids this by choosing small-angle combinations preferentially.
Then the two real jets in the final state of the tube model
are built up from the high-rapidity ends of the tube, and
the first junk-jet to be resolved will be the wide-angle
hadron or cluster with the largest transverse momentum,
\beq
\VEV{y_3}^A\sim\frac{\VEV{p_{t,\mx}^2}}{Q^2}\;,
\eeq
as represented schematically in Fig.~\ref{fig_tubes}(c).
The relationship between $\VEV{p_{t,\mx}^2}$ and $\VEV{p_t}^2$
depends on the multiplicity $N$ and the form of the $p_t$-distribution.
For an exponential $p_t$-distribution we have
\beq\label{ptmax2}
\VEV{p_{t,\mx}^2}\sim\left(\VEV{p_t}\ln N\right)^2
\eeq
and therefore
\beq\label{yAtube}
\VEV{y_3}^A\sim\left(\frac{\lambda}{Q}\ln\ln(Q/\lambda)\right)^2\;.
\eeq
This is shown by the dot-dashed curve in Fig.~\ref{fig_tube3},
which again agrees quite well with the Monte Carlo data (stars).

\subsection{Mis-clustering}
It is difficult to see how the value of $\VEV{y_3}$ could be
reduced much further by refinement of the A algorithm, since
fluctuations in $p_t$ will always lead to a multiplicity-dependent
enhancement of $\VEV{p_{t,\mx}}$ relative to the intrinsic
non-perturbative scale set by $\VEV{p_t}_\NP$. There is however
a remaining weakness of the algorithm that, while
having no perceptible effect on $\VEV{y_3}$, can be expected to
show up more in studies of multi-jet rates and the internal structure
of jets. It has to do with the wrong assignment of soft wide-angle
radiation to {\em already resolved}\/ jets.

Consider the parton-level application of the algorithm to the basic splitting 
$q_1\to q_1+g_2$ accompanied by a soft large-angle gluon $g_3$
(Fig.~\ref{fig_branching}), such that 
$$
E_1\gg E_2\gg E_3\>, \qquad \theta_{12}\ll \theta_{13}\approx\theta_{23}\>.
$$

\FIGURE[htb]{
\epsfig{file=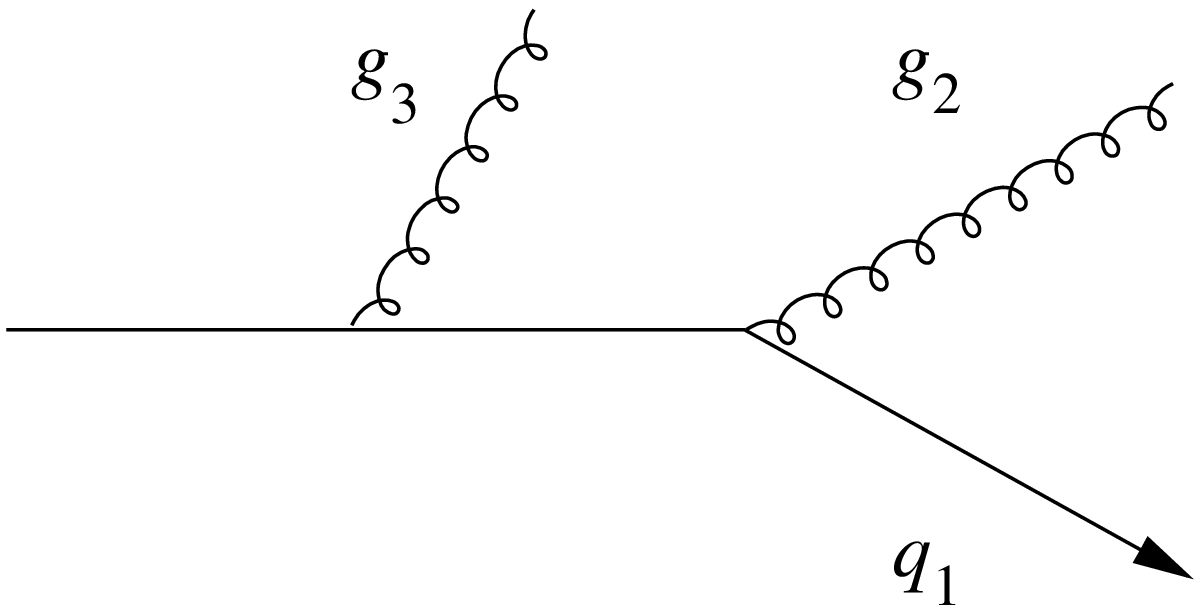,width=9cm}
\caption{Parton branching with soft, large-angle gluon emission.}
\label{fig_branching}
}

According to the angular ordering property of QCD, the soft gluon $g_3$ is
radiated  {\em coherently}\/ and the corresponding
radiation intensity is proportional 
to the total colour charge of the relatively narrow system $q_1+g_2$, which
is just that of the initial quark.  However, in {\em half}\/ of the events
the A (or D) algorithm will erroneously assign $g_3$ to the gluon jet $g_2$
because the latter happens to lie a little closer in angle to $g_3$
than the quark.

At first sight, this does not seem to do any harm. The three partons
are going to be identified as a 1-, 2-, or 3-jet system depending
on the chosen value of $\ycut$, and the corresponding probabilities
of the 1-, 2-, 3-jet configurations remain unaffected by mis-clustering.   
This is true as long as one is concerned only with the jet fractions
and not with the internal structure of the jets.
However, after having resolved this system as {\em two}\/ jets at some
resolution  $y_1$ one may attempt to choose another yet smaller $y_2\ll y_1$
and to study the internal sub-jet structure by examining the
history of merging. In this case the results will be misleading
because one will find that the gluon sub-jet is artificially overpopulated.
The same mis-clustering will occur when all three partons in our example are
gluons, although in this case the practical damage may be less pronounced.

\subsection{Cambridge (C) algorithm}
The modification we propose in order to reduce the mis-clustering
effect consists of
`freezing' the {\em softer} of two resolved objects to prevent it from
attracting any extra partners from among the remaining objects with larger
emission angles. The corresponding {\em Cambridge} (C)
algorithm is thus defined as follows. As before, one starts with a table
of the energies $E_i$ of primary objects and their relative angles as
given by the ordering variable $v_{ij} = 2(1-\cos\theta_{ij})$.

\vspace{1ex}
\noindent{\bf Step 0}. If only one object remains in the table, then store
this as a jet and stop.

\noindent{\bf Step 1}. Otherwise, select the pair of objects ($ij$) having
the minimal value of the ordering variable, $v_{ij}$. Order the
pair such that $E_i\le E_j$.

\noindent{\bf Step 2}. Inspect the value of the test variable,
$$y_{ij} = E_i^2\,v_{ij}\;.$$
\bit
\item If $y_{ij}< \ycut$, then update the table by deleting $i$ and $j$,
introducing a new particle ($ij$) with 4-momentum $p_{ij}=p_i+p_j$, and
recomputing the relevant values of the ordering variable.
\item  If $y_{ij}\ge \ycut$, then store $i$ as a jet and delete it
from the table.
\eit

\noindent{\bf Step 3}. Go to Step 0.

\vspace{1ex}
This procedure suffices to prevent mis-clustering in the dominant
region of phase space where, in the above example, the gluon $g_2$
is softer than the quark $q_1$. It will still cluster the partons
incorrectly in the subleading configuration where the gluon (call it
$g_1$ now since $E_1>E_2$) is accidentally harder than the quark ($q_2$). 
After having resolved them, it will erroneously assign the soft gluon 
$g_3$ to $g_1$ (the more energetic one), %thus confusing
spoiling the association between the clustering and the colour factor.
A similar mistake will occur in the case of a resolved $g\to q_1\bar q_2$
splitting plus a soft wide-angle gluon $g_3$: the algorithm will
inevitably cluster $g_3$  with the quark or the antiquark, which is
incorrect because the source of $g_3$ is the coherent octet colour of
the pair, not a triplet. However, the fraction of events of these
types should be small, and it is hard to see how such occasional
mistakes can be avoided without distinguishing between different
species of partons.

As already mentioned, the additional `soft freezing' step in the
Cambridge algorithm does not affect the quantity $\VEV{y_3}$
significantly. This is shown by Fig.~\ref{fig_tube3}, where
the points for the A and C algorithms coincide within the
plotting resolution.

We do, however, expect the internal jet structure resulting
from the A and C algorithms to be different. The aim of the
freezing is to prevent soft jets from acquiring particles that
do not belong to them. A good diagnostic quantity for
this is the mean number of particles in the first junk-jet
to be resolved in the tube model, i.e.\ the third (softest)
jet when $\ycut=y_3$. In the optimal algorithm this quantity,
$\VEV{n_3}$, should be as small as possible.

\FIGURE[t]{
\epsfig{file=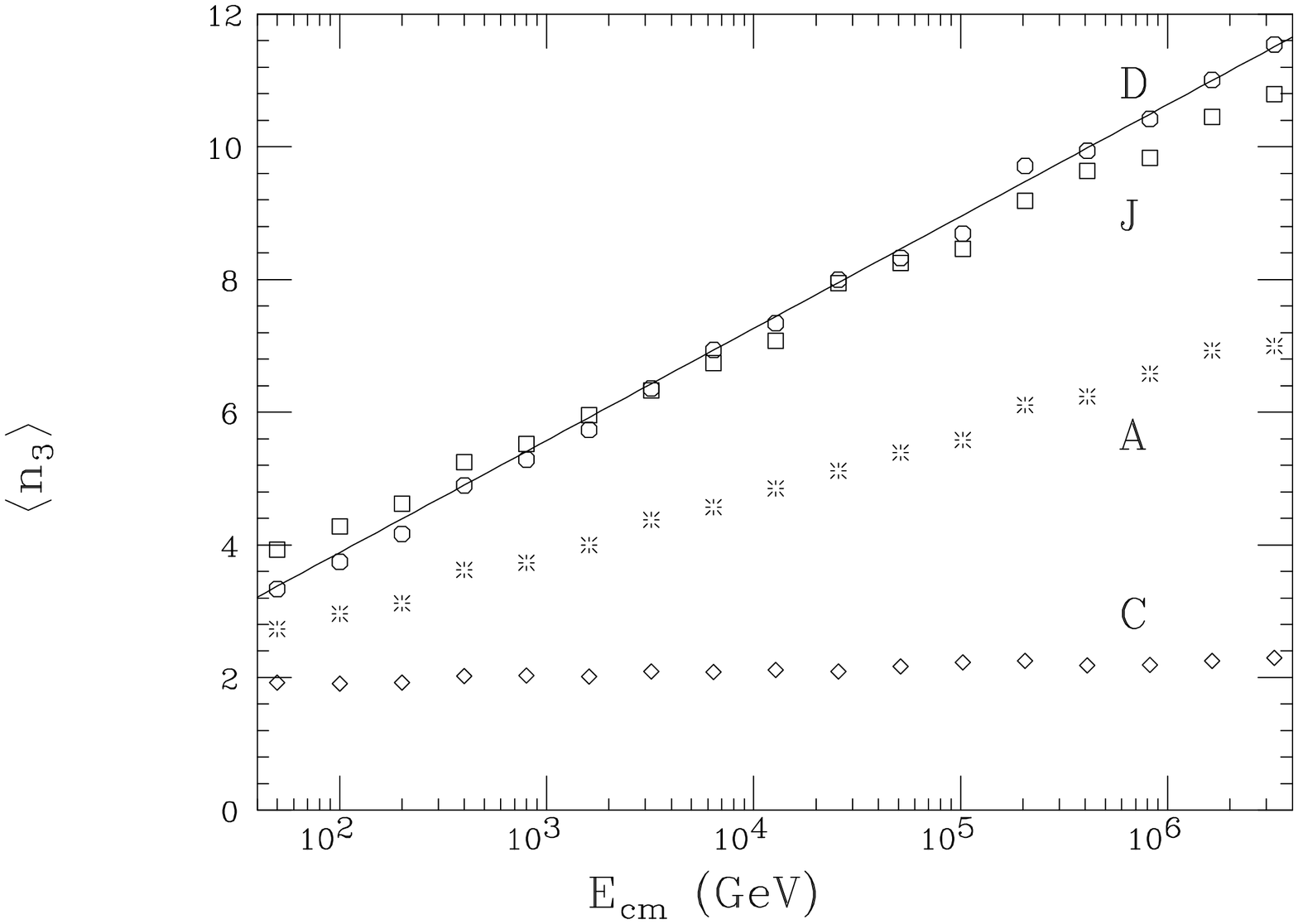,height=16cm,angle=90}
\caption{Mean number of particles in the third jet at $\ycut=y_3$
in the `tube' hadronization model, for the J (square symbols), D (circles),
A (stars) and C (diamonds) jet clustering schemes. The curve
shows $0.22$ times the total number of particles.}
\label{fig_mult3}
}

Our results on $\VEV{n_3}$ for the various algorithms are
presented in Fig.~\ref{fig_mult3}.
As could be guessed from Fig.~\ref{fig_tubes}, the values
in the JADE and original Durham algorithms are quite similar,
both being somewhat below one quarter of the total multiplicity
$N$. As shown by the curve, $\VEV{n_3}^D$ is in fact about $0.22 N$,
with $\VEV{n_3}^J$ rising slightly {\em less} rapidly. Therefore
in this respect the unmodified Durham algorithm performs slightly
{\em worse} than JADE asymptotically.  The angular-ordered
Durham algorithm performs a good deal better, but the population of
the junk-jet still grows steadily with increasing energy. Introducing
the soft freezing step as well (C algorithm) kills the growth of the
spurious jet almost completely, and the value of $\VEV{n_3}^C$
remains near 2 at all energies.

Taking Figs.~\ref{fig_tube3} and \ref{fig_mult3} together, we see that
the Cambridge algorithm performs best in limiting both the amount
of junk-jet formation and the growth of junk-jets once
they have formed.

In real hard processes like $\ee$ annihilation, jet
algorithms have to operate on final states containing both
genuine hard jets and spurious junk-jets, and should ideally
suppress the latter as far as possible while remaining
sensitive to the former. We shall study these issues in
more depth in later Sections, but a few qualitative observations
can be made here. First, we would expect the angular ordering
in the A and C algorithms to reduce the number of jets at a
given value of $\ycut$, relative to the D algorithm, because,
as remarked earlier, separating the ordering and test variables
generally increases the number of combinations attempted.
This effect should become more
pronounced as the number of objects increases, i.e.\ with
increasing energy and/or decreasing $\ycut$. On the other hand,
the soft freezing step in the C algorithm tends to enhance
the number of jets, relative to the A algorithm, by forbidding
combination with frozen jets. Again this difference should be more
pronounced for a larger number of objects. 

It should be emphasised that neither the angular-ordering nor the soft
freezing modifications affect the good properties of the Durham algorithm
with respect to the resummation of large logarithmic terms at small values
of $\ycut$. The leading and next-to-leading logarithms of $\ycut$
come from kinematic regions where the sequence of clustering is
not affected by the extra steps in the modified algorithms.
Thus the only adjustment to resummed predictions will be in the
subleading `remainder' terms, as will be discussed in more
detail in Sec.~\ref{sec:njets}.

In terms of computing, the modified algorithms are slightly
less convenient than the original JADE-type (including Durham)
ones, because of the new distinction between the ordering and
test variables.  In a JADE-type algorithm, the sequence of
clustering is independent of the value of $\ycut$. Therefore
the clustering can be done once and for all, the sequence
can be stored, and all questions about jet multiplicity and
substructure can be answered subsequently without much further
computation. This is the strategy used in the {\tt KTCLUS}
package \cite{KTCLUS}. Once the ordering and test variables
are separated, however, the clustering sequence depends in
general on the value of $\ycut$ and all computations involve
a complete reclustering up to that value.

For example, in the original Durham algorithm the value
of $y_3$ for a given event is defined precisely by the
point in the clustering sequence at which three objects
become two, independently of $\ycut$. In the modified
algorithms, the calculation of $y_3$ involves repeatedly
reclustering the event with different values of $\ycut$,
until the value at which three objects become two is found
with adequate precision, e.g.\ by a binary
search procedure.  The extra computation might have
caused difficulties when the JADE algorithm was originally
proposed, but advances in computing over the past ten
years mean it is no longer a problem.

\section{Fixed-order results}
\label{sec:fixedorder}
It is common in the specialized literature to define the $n$-jet fraction
$f_n(y)$ by\footnote{From now on, in order to simplify the
notation, we shall often use $y$ to represent $\ycut$.} 
\beq
\label{fn}
f_n(y)=\frac{\sigma_n(y)} 
                  {\sum_m \sigma_m(y)}
            =\frac{\sigma_n(y)} 
                  {\st},
\eeq
where $y$ is the jet resolution parameter.
If $\st$ identifies the $total$ hadronic cross section
$\st=\sigma_{0}(1+\as/\pi+ ... )$, $\sigma_{0}$ being the 
lowest-order Born cross section, then the constraint $\sum_n f_n(y)=1$ applies.

For the choice $\mu=Q=\Ecm$ of the renormalization scale, 
one can conveniently write the $n$-jet fractions (e.g., for the cases 
$n=2,3$ and 4) in the following form \cite{BKSS}:
\beq
\label{f2}
f_2(y) = 1 - \left( \frac{\as}{2\pi} \right)    A(y)
           + \left( \frac{\as}{2\pi} \right)^2 (2A(y)-B(y)-C(y)) + ... ,
\eeq
\beq
\label{f3}
f_3(y) =     \left( \frac{\as}{2\pi} \right)    A(y)
           + \left( \frac{\as}{2\pi} \right)^2 (B(y)-2A(y)) + ... ,
\eeq
\beq
\label{f4}
f_4(y) =     \left( \frac{\as}{2\pi} \right)^2  C(y) + ... ,
\eeq
where the coupling constant $\as$ and the functions $A(y), B(y)$ 
and $C(y)$ are defined in some 
renormalization scheme (in what follows we shall make use of the 
$\overline{\mbox{MS}}$ scheme). The terms of order $\cO{\as^2}$
involving $A(y)$ take account of the normalization to $\st$ rather
than to $\sigma_{0}$.

\FIGURE[t]{
\epsfig{file=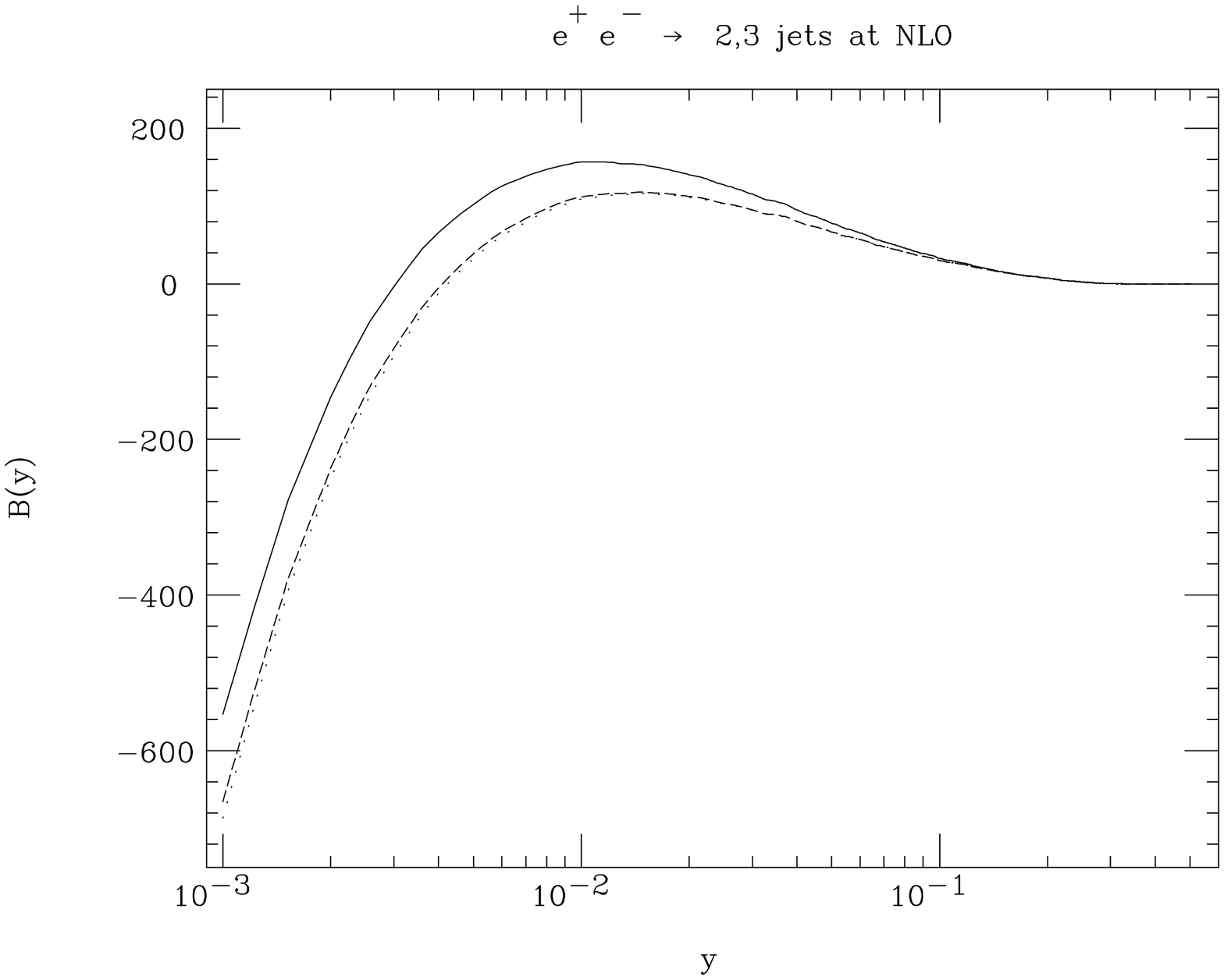,height=16cm,width=12cm,angle=90}
\caption{The parton level $B(y)$ function entering in the two- and three-jet 
fractions (\ref{f2})--(\ref{f3}) at NLO in the D (continuous line), A (dashed
line) and C (dotted line) schemes.}
\label{fig_bfunction}
}

For the original Durham jet algorithm, several studies were carried out
in Ref.~\cite{BKSS}, together with a parametrization of the auxiliary 
functions $A(y), B(y)$ and $C(y)$. There, analytic fits of the form
\beq
\label{fit}
F(y)=\sum_{n=0}^{4} k_n\left( \ln\frac{1}{y} \right)^n
\eeq
were performed for $F=A,B$ and $C$. The coefficients $k_n$ were tabulated
for values of $y$ in the range 0.01--0.33 in the case of $A(y)$ and
$B(y)$, whereas for $C(y)$ the interval was 0.01--0.1.

In this Section we present results similar to those of Ref.~\cite{BKSS}
for the case of the two modifications of the original Durham 
clustering scheme that we are proposing, the `angular-ordered 
Durham' and `Cambridge' schemes, and we compare the
performances of all three algorithms in predicting several results of 
phenomenological relevance in QCD analyses. In particular, in order to
match the advances on the experimental side in resolving jets at very
small values of $y$, in our studies we have extended the range of
the jet resolution parameter down to the minimum figure of $0.001$.
However, in this respect, we have to stress that
at very small values of $y$ (i.e., around 0.001) the perturbative 
calculations become unreliable, because of the growing five-jet contribution, 
which of course does not appear in our $\cO{\as^2}$ computation.  
In other words, for small $y$'s, terms of the form $\cO{\as^n
\ln^m y}$ with $m\le 2n$ become large and have to be resummed 
\cite{y3resum}\footnote{We look forward to the availability in numerical
form (see, e.g., Ref.~\cite{numa3}) of the complete
$\cO{\as^3}$ results, toward which contributions
have been made by two different groups \cite{slac,durham}.}.
In producing the numerical results presented in this Section
we have made use of the Monte Carlo (MC) program {EERAD} \cite{EERAD}, 
which implements the complete next-to-leading-order (NLO) corrections
to the $e^+e^-\rightarrow$ 2 jet and $e^+e^-\rightarrow$ 3 jet rates, as well
as the leading-order (LO) contributions to $e^+e^-\rightarrow$ 4 jets.
For the purpose of illustration, the total c.m.\ energy $Q$
is always chosen to be the mass of the $Z$-boson and no QED 
initial-state-radiation (ISR) has been included. The numerical values
are $ M_Z=91.17$ GeV for the mass and $\Gamma_Z=2.516$ GeV for the
width (as we use the $e^+e^-\rightarrow Z,\gamma^*\rightarrow\dots$ 
annihilation rates).

\FIGURE[t]{
\epsfig{file=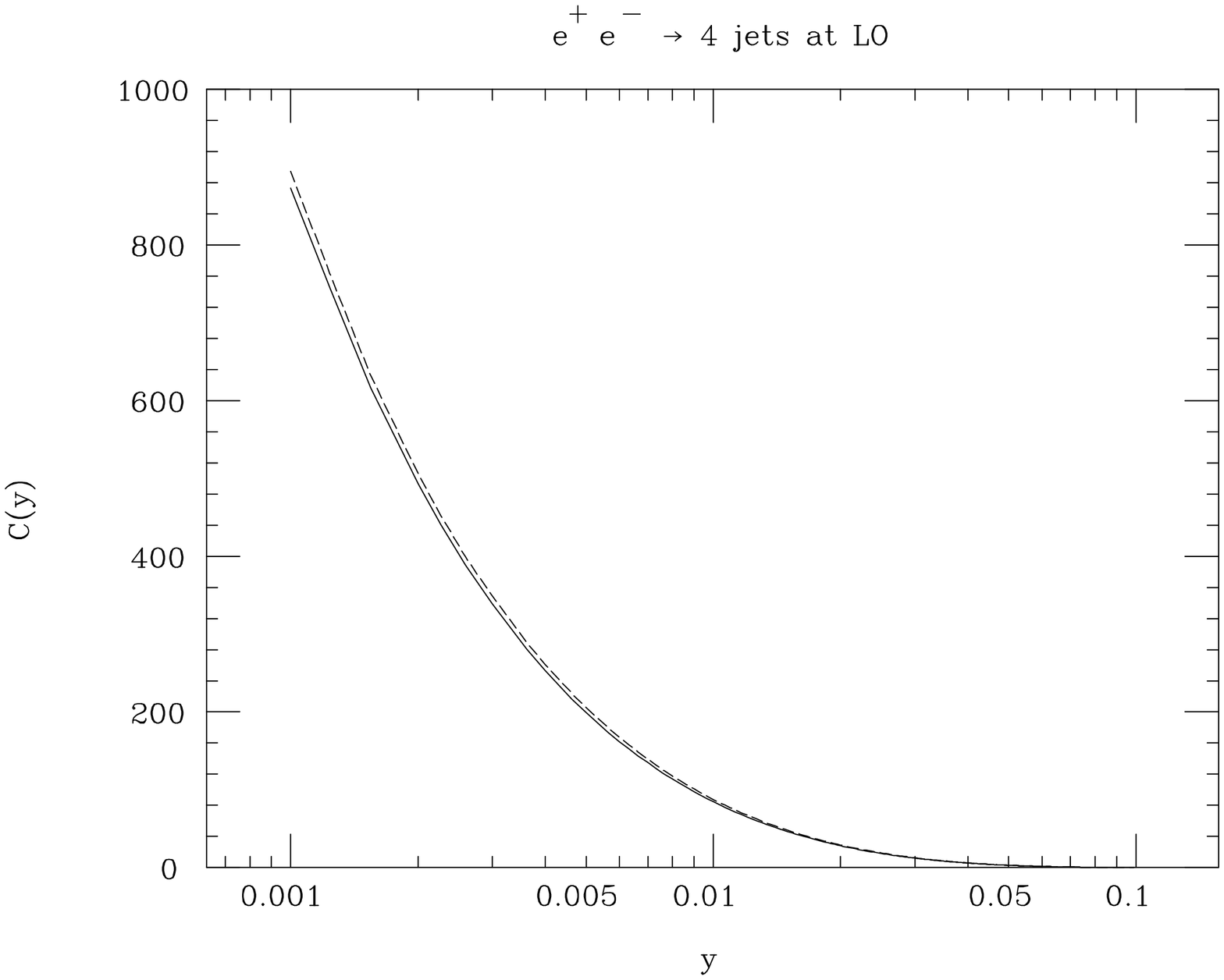,height=16cm,width=12cm,angle=90}
\caption{The parton level $C(y)$ function entering in the four-jet fraction 
(\ref{f4}) at LO in the D, A (continuous line) and C (dashed line) 
schemes.}
\label{fig_cfunction}
}

Up to the order $\cO{\as^2}$, differences in the jet fractions 
(\ref{f2})--(\ref{f4}) as predicted by the three jet algorithms D,
A and C can only occur when calculating the NLO contributions
to the two- and three-jet rates and the LO ones to the
four-jet cross section. More precisely, only the functions $B(y)$ and $C(y)$ 
can vary depending on the scheme adopted. In particular, $B(y)$ turns
out to be different for all three jet algorithms, whereas $C(y)$ is
the same for the D and A schemes but differs for the C scheme.
This can easily be understood if one recalls that, in comparison
to the D algorithm, the A algorithm only involves
a modification in the combination scheme (i.e., the `angular ordering')
used in the $n$-parton $\rightarrow$ $(n-m)$-jet transitions (with $n\ge4$
and $1\le m\le n-1$), whereas the C algorithm also affects the 
$n$-parton $\rightarrow$ $n$-jet rates for $n\ge4$ (i.e., via the
`soft freezing' procedure). Indeed, note that for $n=3$ partons, 
kinematical constraints impose that, on the one hand, the two closest 
particles are also those for which $y$ is minimal and, on the other hand,
the identification of the softest of the three partons as a jet implies that
the remaining two particles are naturally the most energetic and far 
apart. This clearly implies that the $A(y)$ function is the same for 
all three algorithms.  
For the above reasons then, we only plot the distributions for the $B(y)$ and
$C(y)$ functions. This is done in Figs.~\ref{fig_bfunction} and 
\ref{fig_cfunction}, respectively. 

\FIGURE[t]{
\epsfig{file=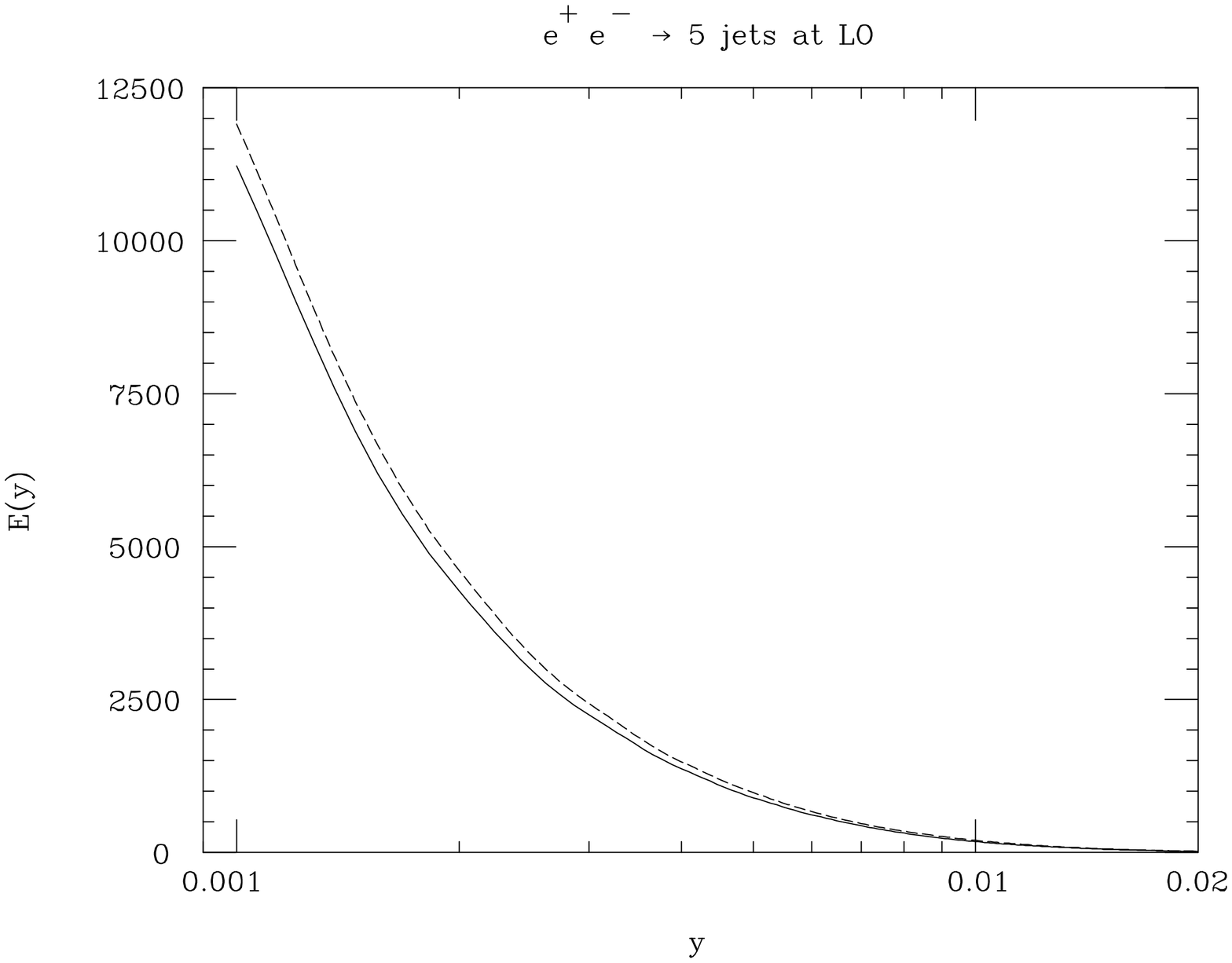,height=16cm,width=12cm,angle=90}
\caption{The parton level $E(y)$ function entering in the five-jet fraction 
at LO in the D, A (continuous line) and C (dashed line) schemes.}
\label{fig_jet5}
}

\FIGURE[t]{
\epsfig{file=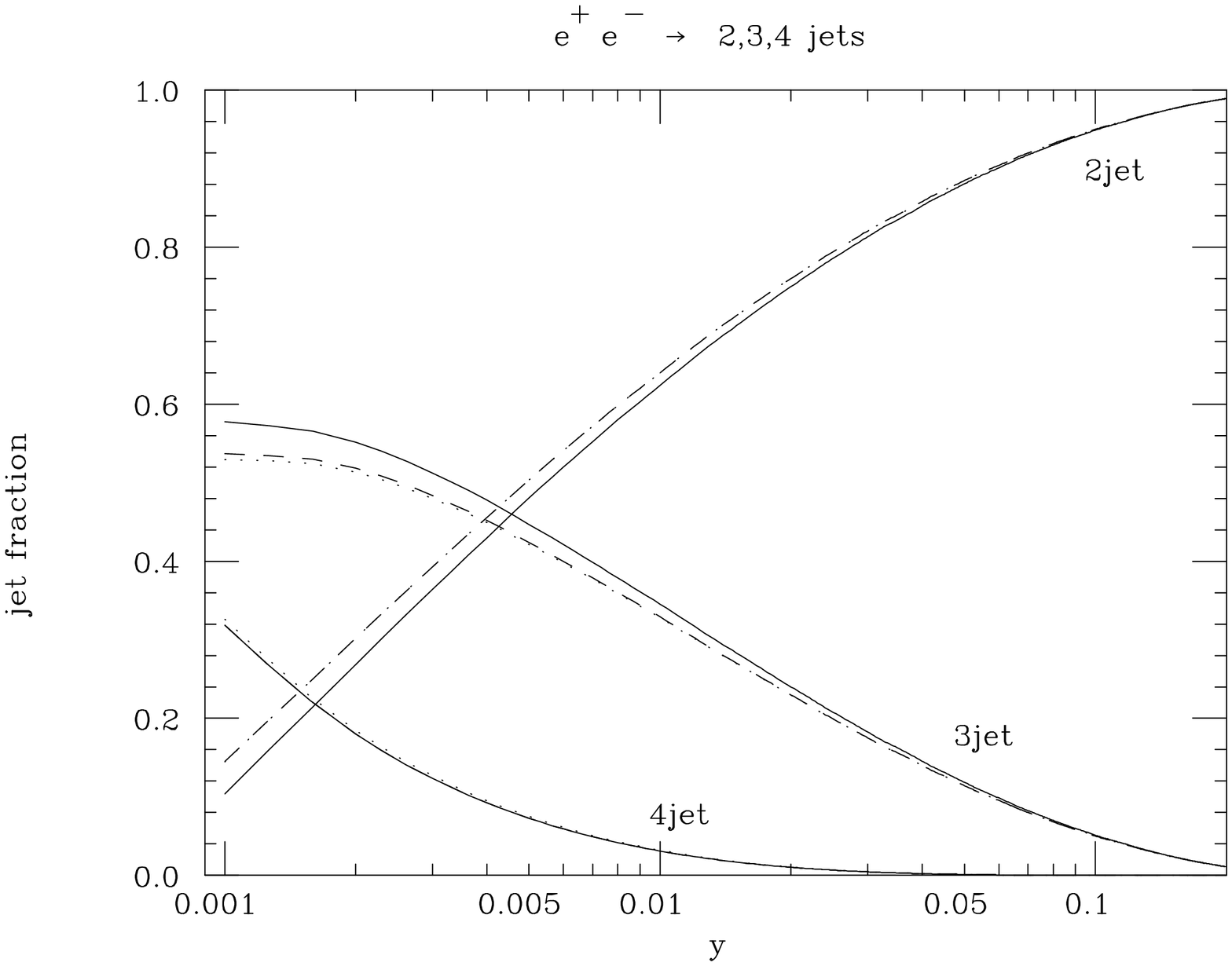,height=16cm,width=12cm,angle=90}
\caption{The $\cO{\as^2}$ two-, three- and four-jet fractions
at parton level in the D (continuous lines), A (dashed lines) and
C (dotted lines) schemes.}
\label{fig_jetrates}
}

Whereas at LO all the contributions to the three-jet cross section
come from the parton process $e^+e^-\rightarrow q\bar q g$,
at NLO contributions to the three-jet rates are due to two sources. First,
the real emission diagrams for the 
processes $e^+e^-\rightarrow q\bar q gg$ and 
$e^+e^-\rightarrow q\bar q Q\bar Q$, in which
one of the partons is unresolved. This can happen when one has either
two collinear partons within one jet or one soft parton outside the
jet cones. Both these contributions are (in general, positively) divergent. 
Thanks to the Bloch-Nordsieck \cite{BN} and Kinoshita-Lee-Nauenberg
\cite{KLN} theorems (see also Ref.~\cite{Sterman}), these collinear
and soft singularities are cancelled at the same order in $\as$  by the 
divergent contributions (generally negative) provided by the second source,
namely the virtual loop graphs. Therefore, after renormalizing
the coupling constant $\as$ a finite three-jet cross section is obtained.
The function $B(y)$ accounts for the above-mentioned three- and four-parton 
contributions. The LO four-jet cross section at a given $y$-value 
is due to four-parton events for which all the measures $y_{ij}$ 
are greater than the resolution parameter. These contributions are contained
in the $C(y)$ term. 

From Fig.~\ref{fig_bfunction}, one can appreciate that,
whereas differences between the two proposed 
modifications of the Durham algorithm are negligible, those
between these two and the original one can amount to several tens of percent
in a sizable part of the experimentally relevant $y$-range, especially for
$y\Ord0.01$. Let us then investigate their behaviour.
It is clear that $B(y)$ 
for the A and C algorithms is consistently smaller (both in
magnitude and sign) than for the D one. This turns out to produce a smaller
three-jet fraction, compare to Eq.~(\ref{f3}), where the $B(y)$ term enters
with a positive sign (the leading piece proportional to
$A(y)$ clearly dominates). This is rather natural since, as discussed
in Sect.~\ref{subsec:Aalg}, the D algorithm represents a sort
of upper bound in jet multiplicity to its variation A. 
The `angular ordering' procedure tends to combine more pairs
of partons than the original D clustering does, as the former would
pair the two closest particles even when their $y$-value is not the 
smallest of the entire event. If these two partons yield
the minimum $y$, then the two combination schemes coincide.
Conversely, this inevitably produces an enhancement of the lowest order
jet fraction, that is the two-jet one, see Eq.~(\ref{f2}), where $B(y)$ is 
indeed preceded 
by a minus sign and the $C(y)$ term is unaffected by the A prescriptions. 
The additional step of `soft freezing' implemented in the C algorithm
tends to enhance the final jet multiplicity
of the original D scheme, by preventing the softer particle in a
resolved jet pair from attracting the remaining 
particles (at large angle) into unresolved parton pairs, 
which would then be merged together, producing
a lower number of final jets. In this respect, therefore, 
the yield of the clustering procedure of the original D algorithm acts,
in terms of final jet multiplicity, as a 
lower bound on that produced by the implementation of the `soft freezing' step.
This can be seen from Fig.~\ref{fig_cfunction}, 
for the auxiliary function $C(y)$ to
which the four-jet rate (\ref{f4}) is proportional, though the effect remains 
at the level of $2-4\%$ over the considered range in $y$.
The increase of the four-jet rate is compensated for by a further decrease 
(with respect to the A scheme)
of the C-scheme three-jet fraction, as can be seen in the $B(y)$ term 
from Fig.~\ref{fig_bfunction} 
and Eq.~(\ref{f3}). However, contrary to the `angular ordering' 
case, when `soft freezing' is adopted the two-jet rates are much less affected,
as both the two terms $B(y)$ and $C(y)$ vary and in opposite directions.

As an additional exercise, in order to show that, as previously discussed,
the effects of `soft freezing' do become more evident as the final state 
multiplicity increases, 
we plot in Fig.~\ref{fig_jet5} the $(\as/2\pi)^3$ coefficient of
the five-jet fraction at LO (which we call $E(y)$), as obtained from the 
programs already employed in Refs.~\cite{5j,moriond}. Here, differences
between, on the one hand, the D and A schemes and, on the other hand,
the C one, can be quite large, more than twice those obtained in the case of
the $C(y)$ function (compare to Fig.~\ref{fig_cfunction}). For example, at
$y=0.001$ these amount to $6\%$ and become relatively larger with $y$. At
the upper end of the interval considered in Fig.~\ref{fig_jet5}, 
$y=0.02$, they are at the level of $\approx 15\%$. 

\FIGURE[t]{
\epsfig{file=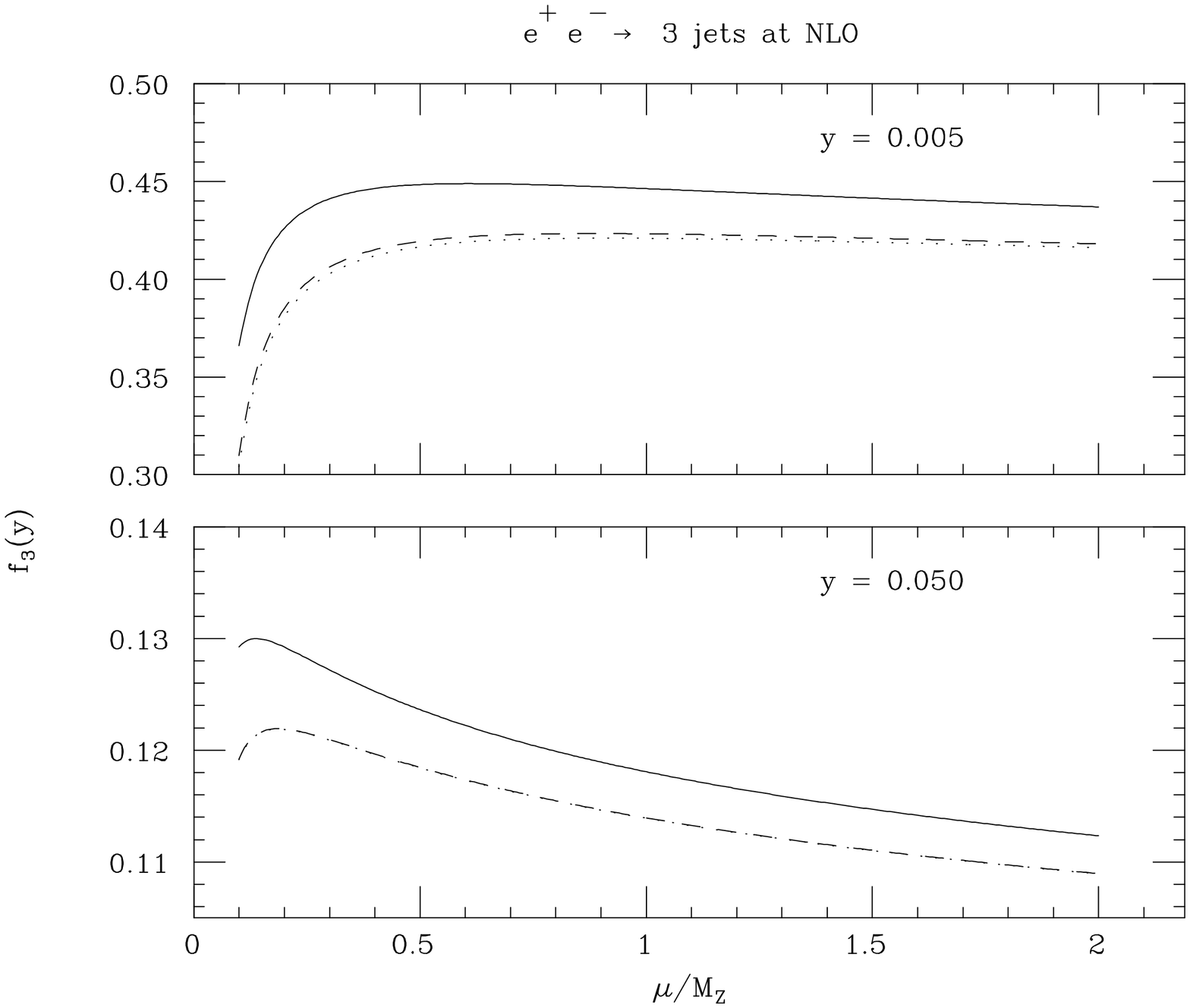,height=16cm,width=12cm,angle=90}
\caption{The parton level three-jet fraction at NLO as a function of the 
renormalization scale $\mu$ at $y=0.005$ and $y=0.050$, for  
the D (continuous lines), A (dashed lines) and C (dotted lines) schemes.}
\label{fig_scale}
}

In summary, at the order
$\cO{\as^2}$, one could conclude the following: whereas 
`angular ordering' tends to shift a part of the three-jet rate into
the two-jet fraction without affecting the four-jet cross section, 
`soft freezing' enhances the fraction of four-jet events and reduces
the three-jet one, with a small effect on the two-jet rate. Note that 
the A scheme only implements the first step, so that the corresponding effect
is more promptly visible. In contrast, as the C scheme adopts both these 
procedures, in this case the two effects tend to mix together in the 
final rates. However, in this case one expects the `angular ordering' effects 
to be dominant over those due to the `soft freezing', as can be argued by 
comparing Figs.~\ref{fig_bfunction} and \ref{fig_cfunction}.
These effects are summarised in Fig.~\ref{fig_jetrates}, where the
two-, three- and four-jet fractions as defined via
eqs.~(\ref{f2})--(\ref{f4}) are plotted. 
As we do not possess the complete $\cO{\as^3}$  result, note that we are
unable to present similar results for the five-jet fraction.
The value adopted in Fig.~\ref{fig_jetrates} for $\as$ is 0.120, 
as given in Ref.~\cite{as}. 

In the remainder of this Section we focus our attention on the three-jet
fraction. In particular, we would like to point out that, since 
the algorithms D, A and C have identical three-jet rates at LO, it is 
evident from Fig.~\ref{fig_jetrates} 
and Eq.~(\ref{f3}) that the NLO corrections to $f_3(y)$ contained in the 
coefficient function $B(y)$ are significantly smaller in the two new 
algorithms. 
From the point of view of the convergence of the QCD asymptotic expansion,
and hence of the reliability of the theoretical predictions,  
this should represent an improvement. In the sense that, provided one
adopts a suitable value of the scale parameter $\mu$ regulating the 
corrections (see Ref.~\cite{subtraction} for discussions of
the best choice of the subtraction scale), it might not be unreasonable 
to argue that jet algorithms having smaller NLO terms may also have smaller
higher-order corrections. Clearly, this can be assessed only when the latter
are computed. However, some hints on this point can
be deduced from studying the scale dependence of the complete NLO results.
In fact, we know that the $\cO{\as^3}$ corrections are guaranteed
to cancel the $\mu$-dependence of the $\cO{\as^2}$ three-jet fraction
up to the order $\cO{\as^4}$. 

The $\mu$-dependence of the three-jet rate is simply introduced in
Eq.~(\ref{f3}) by the two substitutions
\beq
\label{scaledep}
\as\rightarrow \as(\mu),\qquad\qquad\qquad 
B(y)\rightarrow B(y)-A(y)\beta_0\ln\frac{Q}{\mu},
\eeq
where $\beta_0$ is the first coefficient of the QCD $\beta$-function,
that is $\beta_0=11-2N_f/3$, with $N_f$ the number of flavours active
at the energy scale $\mu$. Our results are shown in Fig.~\ref{fig_scale},
where the value of $f_3(y)$ is plotted as a function of the scale parameter
$\mu/Q\equiv \mu/M_Z$, over the range between $1/10$ and
$2$, at two fixed values of the jet resolution parameter. Note that 
for the strong coupling constant we have used the two-loop expression, 
with $N_f=5$ active flavours and 
$\Lambda^{(5)}_{\overline{\mbox{\tiny{MS}}}}=250$ 
MeV (yielding $\as(M_Z)=0.120$, to match the rates given in 
Fig.~\ref{fig_jetrates}). 

Although the structure of the QCD perturbative expansion does not prescribe 
which value should be adopted for the scale $\mu$, an obvious requirement
is that it should be of the order of the energy scale involved in the problem 
(i.e., the c.m.\ energy $Q$). Such a choice prevents the 
appearance of large terms of the form $(\as\ln(\mu/Q))^n$
in the perturbative series. Therefore, it is reassuring to recognise that
in Fig.~\ref{fig_scale} the A and C curves are more stable than
the D ones around the value $\mu/Q=1$, especially at smaller $y$-values,
where the differences in $f_3(y)$ among the three schemes are larger.
For example, over the range $1/2\leq \mu/Q \leq 2$
the variation of $f_3(y)$ between its maximum and minimum values at
$y=0.005$ is $\approx2.7\%$ for the D scheme and $\approx1.3\%$ for A
and C. For $y=0.050$ differences between the three algorithms 
are smaller, as $f_3(y)$ varies by $\approx10\%$ in the first scheme
and $\approx8.7\%$ in the other two. 

\TABLE[t]{
\begin{tabular}{|c|c|c|c|c|c|c|c|}
\hline
\rule[0cm]{0cm}{0cm}
$F$ & Algorithm & $y$-range & $k_0$ & $k_1$ & $k_2$ & $k_3$ & $k_4$ 
                      \\ \hline\hline
\rule[0cm]{0cm}{0cm}
$A$ & D, A, C & $0.001-0.20$ & 
    $0.843$ & $-2.177$ & $1.237$ & $-0.0287$ & $0.00312$ \\ 
\hline
\rule[0cm]{0cm}{0cm}
$B$ & D  & $0.001-0.20$ & 
    $49.287$ & $-64.104$ & $14.639$ & $7.955$ & $-1.529$ \\ 
$B$ & A & $0.001-0.20$ & 
    $26.869$ & $-33.941$ & $2.930$ & $9.053$ & $-1.573$  \\ 
$B$ & C & $0.001-0.20$ & 
    $28.970$ & $-37.056$ & $4.448$ & $8.800$ & $-1.569$  \\ 
\hline
\rule[0cm]{0cm}{0cm}
$C$ & D,A & $0.001-0.10$ &
    $2.370$ & $-12.736$ & $15.431$ & $-7.070$ & $1.121$  \\ 
$C$ & C   & $0.001-0.10$ & 
    $-1.405$ & $-7.776$ & $13.201$ & $-6.705$ & $1.111$  \\ 
\hline
\end{tabular}
\caption{Parametrization of the three- and four-jet QCD functions
$A$, $B$ and $C$ as polynomials $\sum_n k_n(\ln(1/y))^n$, for the
Durham algorithm and its variants. The range of validity in $y$
is given for each case.}
\label{tab:kparam}
}

We conclude this Section by presenting a polynomial fit of
the form (\ref{fit}) to the $A$, $B$ and $C$ functions in the D, A and C
schemes. The range of validity of our parametrization extends from 
$y=0.001$ (for all three auxiliary functions) up to $y=0.2$ for $A$ and $B$ 
and $y=0.1$ for $C$. This should allow for an easy comparison between data
and perturbative predictions over the
$y$-range normally adopted in phenomenological studies (see, e.g., 
Ref.~\cite{OPALnj}). The values of the coefficients $k_n$, with $n=0, ... 4$,
are reported in Tab.~\ref{tab:kparam}.

\section{Mean number of jets}
\label{sec:njets}
In Sect.~\ref{sec:algo} we studied the mean value of $y_3$,
the value of $\ycut$ at which three jets are just resolved,
as a measure of jet algorithm performance, using a simple
`tube' model to estimate hadronization effects. That model had
no perturbative contribution to $y_3$, and so we could explore
the region of low $\ycut$ simply by increasing
the energy (see Fig.~\ref{fig_tube3}).
In the presence of perturbative contributions, $\VEV{y_3}$ is not
so small (about 0.02 at $Q=M_Z$). Therefore for more realistic
investigation of the interplay between perturbative and non-perturbative
effects, we choose here to investigate a different quantity, namely the
mean number of jets, $n_{\mbox{\tiny jets}}$.
This may be written in terms of the individual 
$n$-jet fractions (\ref{fn}) as
\begin{equation}\label{n_jets}
n_{\mbox{\tiny jets}}\equiv\cN(y)=\sum_{n=1}^\infty n f_n(y).
\end{equation}

The mean number of jets has the advantage that it can be
studied as a function of the jet resolution $y$,
down to arbitrarily low
values, at fixed energy. Furthermore, its perturbative
behaviour at very low values of $y$ can be computed
including resummation of leading and next-to-leading
logarithmic terms to all orders \cite{CDFW1}.
This behaviour can be matched with the fixed-order
results of the previous section, to give reliable
predictions (apart from non-perturbative contributions)
throughout the whole range of $\ycut$.  Non-perturbative
contributions can then be estimated
by comparing the perturbative results with those of a
Monte Carlo event generator such as HERWIG \cite{Herwig}.

\subsection{Resummed predictions}
Using the notation of the previous Section, we have from
Eqs.~(\ref{f2})-(\ref{f4})
\beq
\label{Nypert}
\cN(y) = 2 + \left( \frac{\as}{2\pi} \right)    A(y)
           + \left( \frac{\as}{2\pi} \right)^2
             \left( B(y)+2C(y)-2A(y) \right) + ...\;.
\eeq
The behaviour of the first-order coefficient $A(y)$ at small $y$ is
of the form
\beq
A(y)= C_F\left(\ln^2 y + 3\ln y + r(y)\right) \; ,
\eeq
where $C_F=4/3$ and the non-logarithmic contribution is \cite{CDOTW,BS2}
\beq\label{ry}
r(y) = 6\ln 2 + \frac{5}{2} - \frac{\pi^2}{6}
+ 4\left(\ln(1+\sqrt 2)-2\sqrt 2\right)\sqrt y - 3.7 y\ln y +\cO y \; .
\eeq
For the second-order coefficient the expected behaviour, for all
three versions of the algorithm, is
\beq\label{Fy}
F(y)\equiv B(y)+2C(y)-2A(y)= C_F \left[\frac{1}{12}C_A\ln^4 y
- \frac{1}{9}(C_A-N_f)\ln^3 y + \cO{\ln^2 y} \right]\;,
\eeq
where $C_A=3$. The terms of order $\ln^2y$ depend in general on the
version of the algorithm (D, A or C). To find them, we made a fit of
the form (\ref{fit}), restricted now to the region $0.001<y<0.02$, with
the coefficients $k_3$ and $k_4$ fixed at the values prescribed by
Eq.~(\ref{Fy}) and $k_0$, $k_1$, $k_2$ as free parameters. The
results are given in Tab.~\ref{tab:smally}. Using these fits for
$y<0.01$ and those in Tab.~\ref{tab:kparam} for $0.01<y<0.1$,
we obtain simple, smooth parametrizations of the second-order
coefficient $F(y)$ over the whole range of $y<0.1$.

\TABLE[t]{
\begin{tabular}{|c|c|c|c|c|c|}
\hline
\rule[0cm]{0cm}{0cm}
Algorithm & $k_0$ & $k_1$ & $k_2$ & $k_3$ & $k_4$ 
                      \\ \hline\hline
\rule[0cm]{0cm}{0cm}
D  & $58.334$ & $-37.546$ & $13.420$ & $-0.2963$ & $0.3333$ \\ 
A  & $88.817$ & $-45.218$ & $11.554$ & $-0.2963$ & $0.3333$ \\ 
C  & $113.20$ & $-58.455$ & $13.412$ & $-0.2963$ & $0.3333$ \\ 
\hline
\end{tabular}
\caption{Parametrization of the second-order coefficient in the
average number of jets as a polynomial $\sum_n k_n(\ln(1/y))^n$,
for the Durham algorithm and its variants.
The range of validity is $y<0.02$.}
\label{tab:smally}
}

To obtain resummed perturbative predictions for the mean number of
jets, we now proceed as in Ref.~\cite{CDFW1}. To next-to-leading
logarithmic (NLL) accuracy, the resummed results are independent of
the version of the algorithm. Therefore the only modifications to the
predictions for the original D algorithm come from the matching to
the fixed-order results given above. We simply subtract the
first- and second-order terms of the NLL resummed result and
substitute the corresponding exact terms. Denoting by $\cN_q$ the
NLL multiplicity in a quark jet, given in \cite{CDFW1}, we obtain
\beq\label{nfin}
\cN(y) = 2\cN_q(y) + C_F \left(\frac{\as}{2\pi} \right) r(y)
           + \left( \frac{\as}{2\pi} \right)^2
             \left( F(y)-2F_q(y) \right)
\eeq
where $F_q$ is the second-order coefficient in $\cN_q$,
given in \cite{CDW2}: 
\beq
F_q(y) = C_F \left\{\frac{1}{24}C_A\ln^4 y
- \frac{1}{18}(C_A-N_f)\ln^3 y
+ \frac{N_f}{9}\left(1-\frac{C_F}{C_A}\right)
\left[\left(4\frac{C_F}{C_A}-1\right)\frac{N_f}{C_A}
-1\right]\ln^2y \right\}\;.
\eeq

The resulting predictions for the three algorithms,
using $\as=0.12$ as in the previous Section, are shown in
Fig.~\ref{fig_njresum}.
\FIGURE[t]{
\epsfig{file=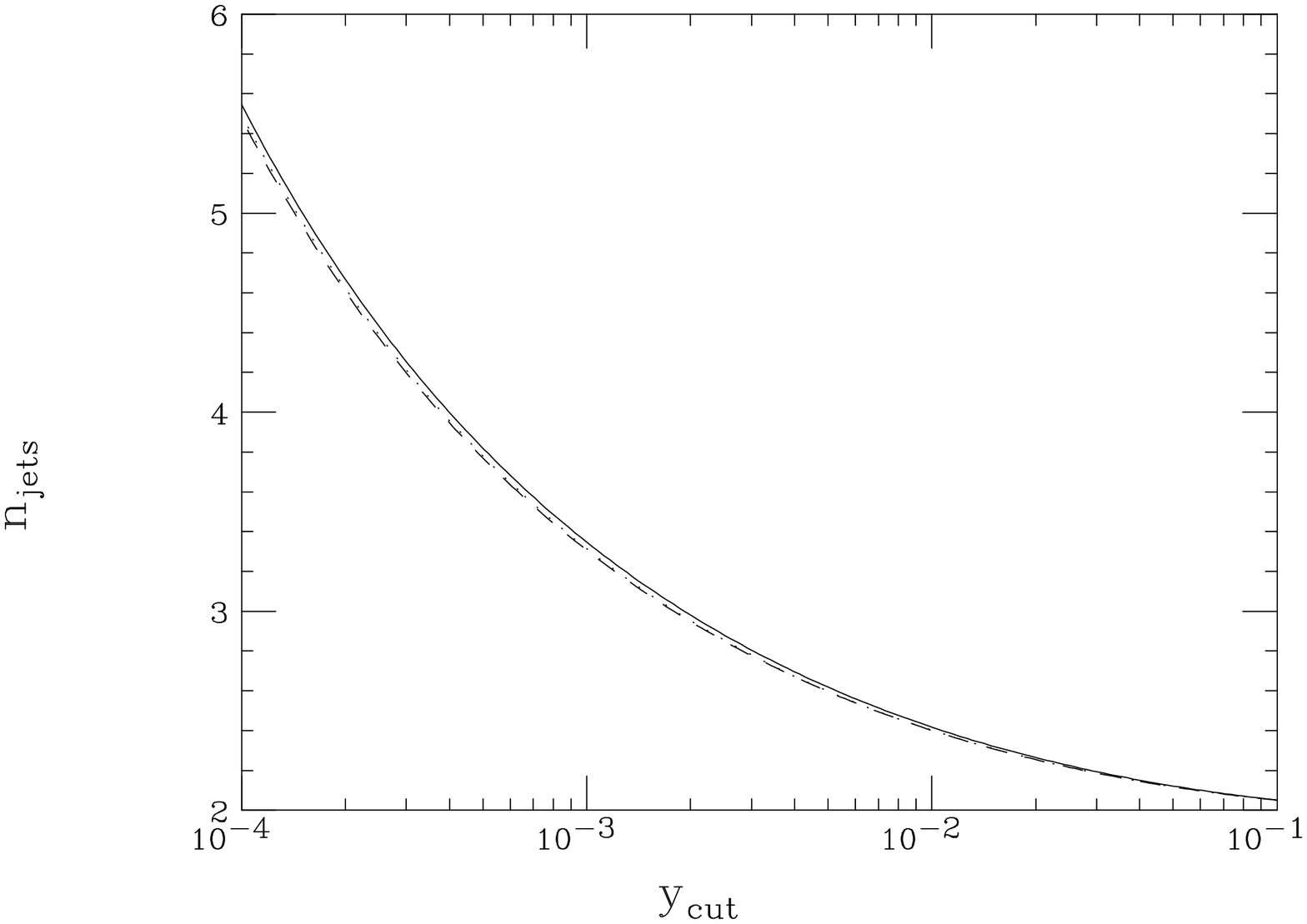,height=16cm,angle=90}
\caption{Resummed predictions for the mean number of jets at parton
level in the D (continuous lines), A (dashed lines) and
C (dotted lines) schemes.}
\label{fig_njresum}
}

\subsection{Monte Carlo studies}
We have used the Monte Carlo event generator HERWIG \cite{Herwig}
(version 5.9 \cite{HW59}) to study non-perturbative corrections to
the above parton-level predictions. The program generates a shower
by coherent branching of the partons involved down to a fixed
transverse momentum scale $\sim 1$ GeV, and then converts these
partons into hadrons using a cluster hadronization model.  The jet
algorithms can be run both on the results of the parton shower and on
the final-state hadrons: the difference between these is customarily
used as an estimate of the hadronization correction.

% The command
%\afterpage{\clearpage}
% should force all current floats to be placed on successive pages immed
% after the end of the current page.
\FIGURE[!p]{
\epsfig{file=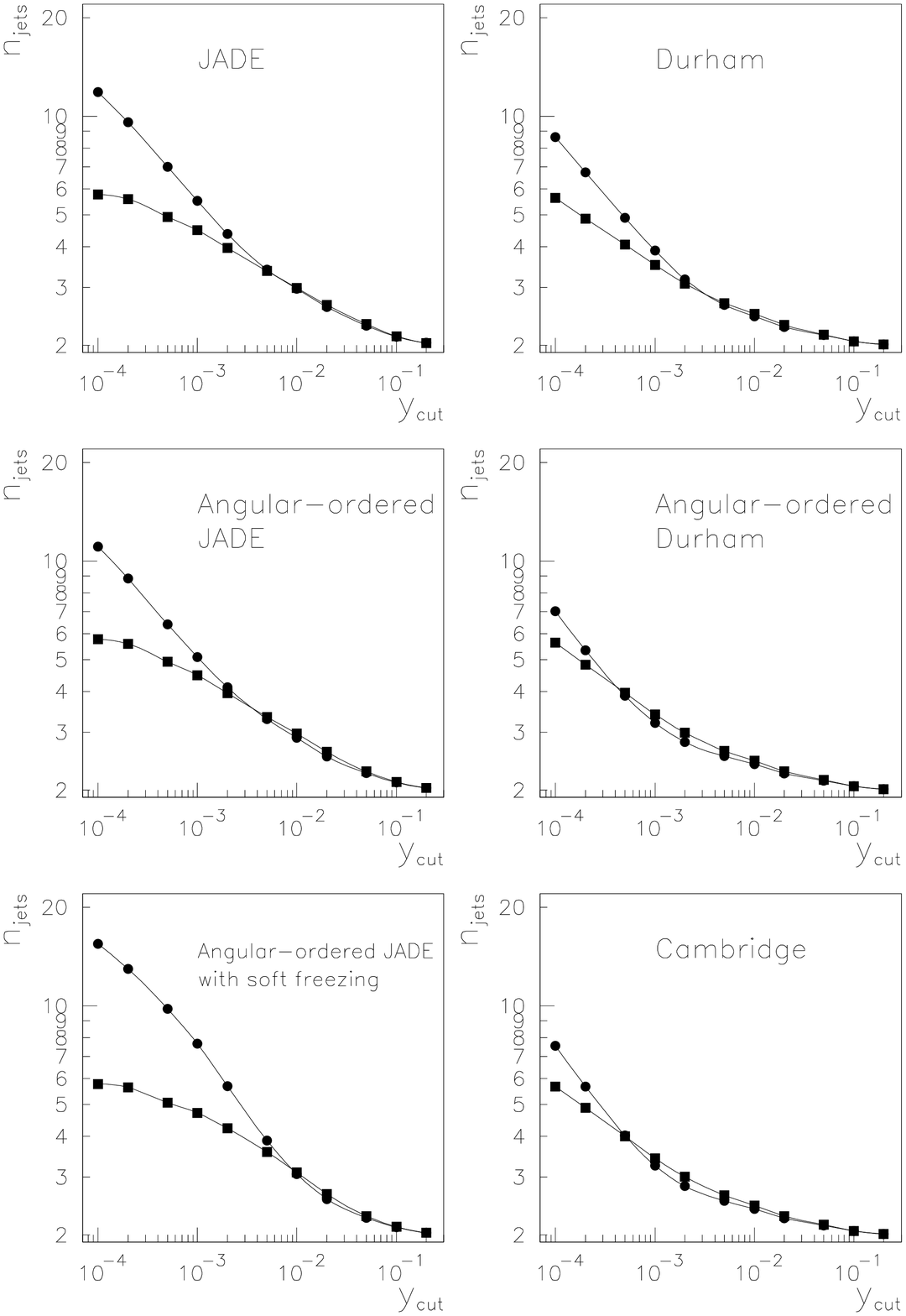,height=20cm}
\caption{Parton-level (squares) and hadron-level (circles) results from
HERWIG on the mean number of jets at $Q=M_Z$
for various jet algorithms as labelled.}
%as a function of the jet resolution variable $\ycut$. The
%statistical errors are smaller than the size of the points.}
\label{fig_njets90}
}

Fig.~\ref{fig_njets90} shows the results for various jet algorithms
at $Q=M_Z$. On the right are the Durham algorithm and the
two variants of it proposed in this paper, with angular ordering
(A) and with angular ordering and soft freezing (C).  On the left,
for comparison, are the JADE algorithm and its corresponding
variants, in which the same ordering and freezing steps are
introduced, but the JADE test variable (\ref{Mijdef}) replaces
(\ref{ktijdef}).

Since the aim was to study the specific process $\ee\to$ hadrons
at a scale equal to the full centre-of-mass energy, HERWIG was set
not to produce any initial-state radiation ({\tt ZMXISR}=0).
Each data point was calculated using 2000 events. A different pair
of random number seeds was used for each value of
$\ycut$ in order for the results to be uncorrelated.
The statistical errors are smaller than the size of the points.

At the parton level, the effects of the modifications to the basic
JADE and Durham algorithms are scarcely visible in
Fig.~\ref{fig_njets90}.
Looking in more detail at the Durham results in comparison with
the resummed predictions (Figs.~\ref{fig_njet_D}-\ref{fig_njet_C}),
we see that in each case HERWIG at the parton level (squares)
somewhat overestimates the number of jets, corresponding to a larger
effective value of $\as$.  The dot-dashed curves correspond to $\as=0.120$,
as in Fig.~\ref{fig_njresum},
whereas the dashed ones, which lie close to the Monte Carlo results,
are for $\as=0.126$. The value $\as=0.114$ (dotted) corresponds
to interpreting the HERWIG input parameter {\tt QCDLAM}, for which
we used the default value of 0.18 GeV, as the NLO scale parameter
$\lms$. Such an interpretation is only justified in a small region
of phase space (see \cite{CMW}) which we would not expect to be
dominant, so a larger effective value is not surprising.

\FIGURE[htb]{
\epsfig{file=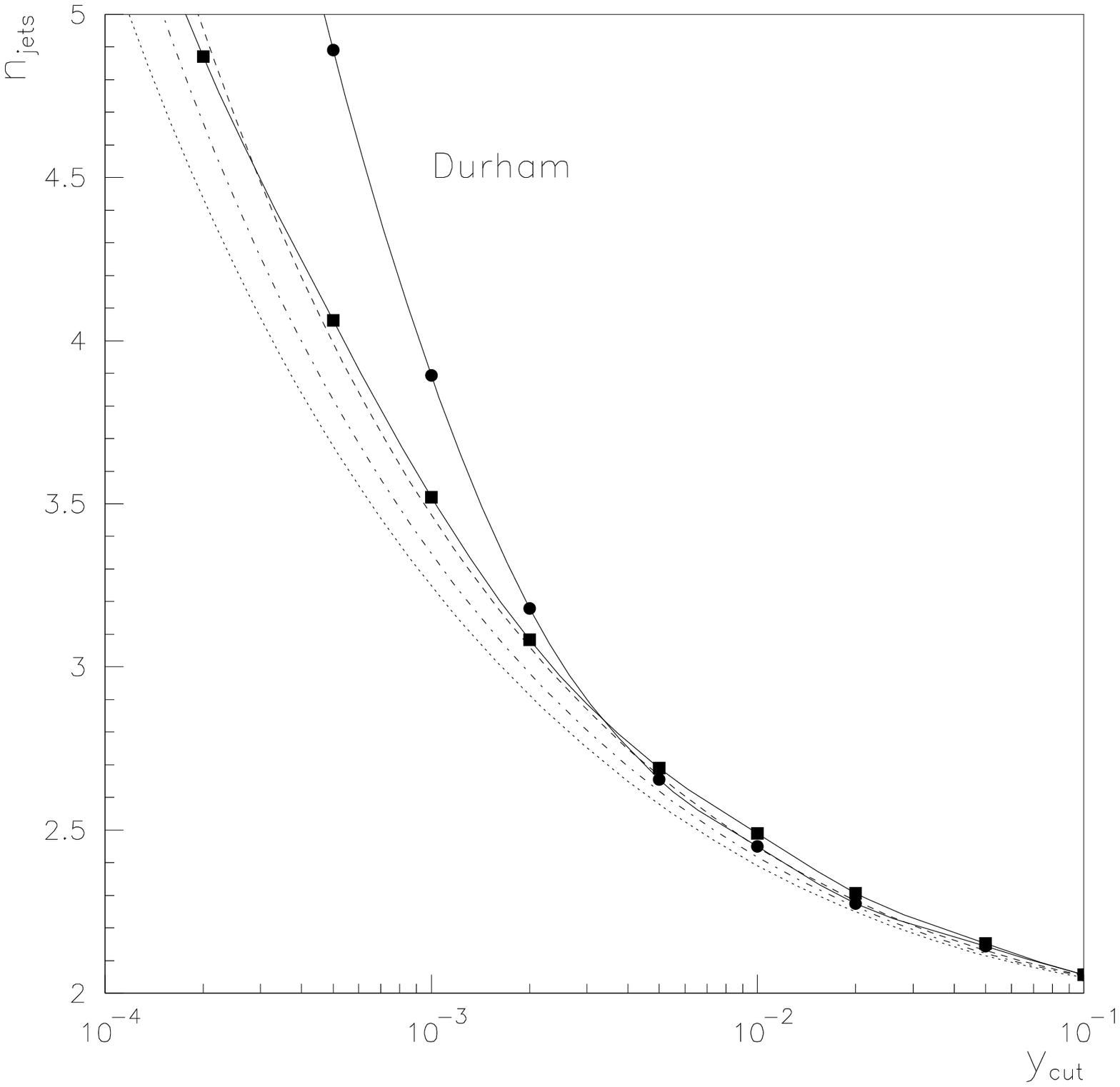,width=12cm}
\caption{Results on the mean number of jets at $Q=M_Z$
for the original Durham algorithm.
HERWIG: squares, parton level; circles, hadron level.
Resummed: dashed, $\as=0.126$; dot-dashed, $\as=0.120$;
dotted, $\as=0.114$.}
\label{fig_njet_D}
}
\FIGURE[htb]{
\epsfig{file=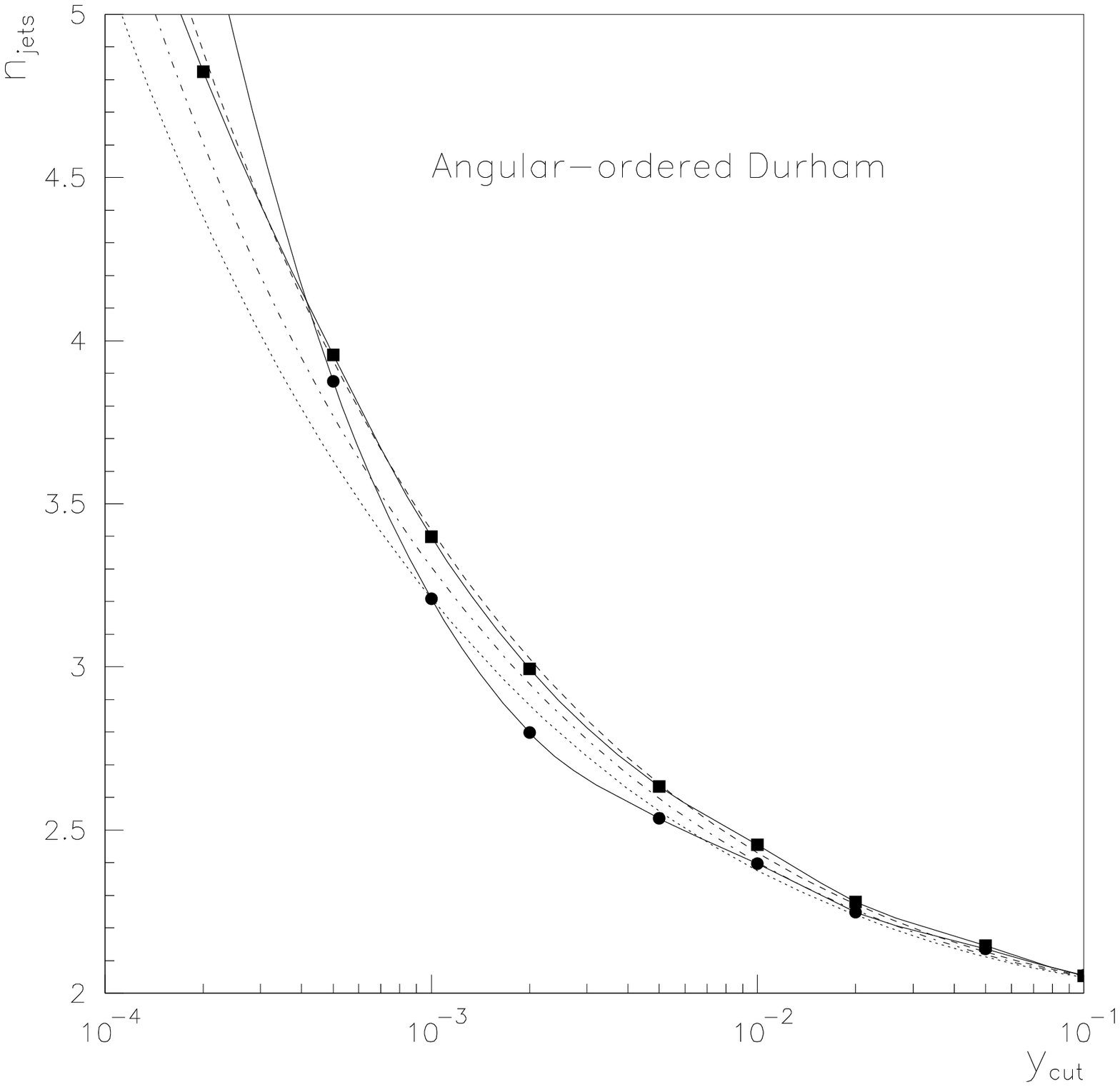,width=12cm}
\caption{Results on the mean number of jets at $Q=M_Z$
for the angular-ordered Durham algorithm. Points and curves
as in Fig.~\ref{fig_njet_D}.}
\label{fig_njet_A}
}
\FIGURE[htb]{
\epsfig{file=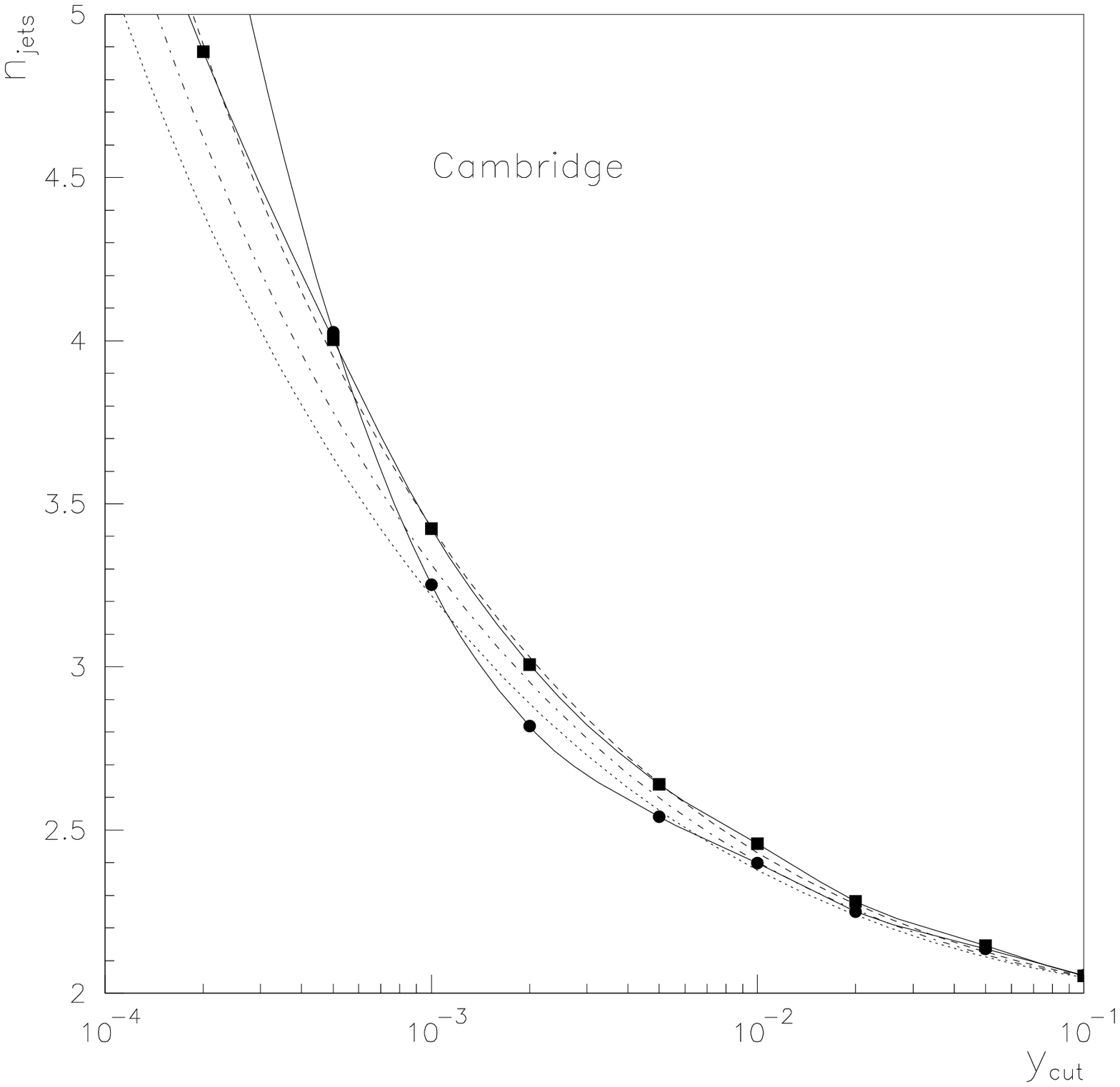,width=12cm}
\caption{Results on the mean number of jets at $Q=M_Z$ for the
Cambridge algorithm. Points and curves as in Fig.~\ref{fig_njet_D}.}
\label{fig_njet_C}
}

Turning to the hadron-level results, we see
that the differences between the various algorithms are much more
evident than at parton level. In particular, the jet multiplicity
is reduced at $\ycut\Ord 10^{-2}$ in the A and C algorithms,
resulting in substantially smaller hadronization corrections   
for $\ycut\Ord 10^{-3}$. At larger values, the hadron-level
points even fall below the parton-level ones, suggesting
a negative hadronization effect or possibly a positive
non-perturbative contribution at the HERWIG parton level,
which could account for the apparently high effective $\as$.
The presence of cut-offs and kinematic boundaries in the
parton shower could indeed give rise to non-perturbative
parton-level contributions. This illustrates the potential
danger of interpreting the hadron--parton difference as an
estimate of non-perturbative effects and simply adding it to
the resummed prediction. In the case of the A and C algorithms,
our results suggest this would lead to an overestimate of $\as$.

Comparing the hadron-level results in Fig.~\ref{fig_njets90},
it can be seen that the angular
ordering criterion alone is not enough to guarantee low hadronization
corrections: it is also necessary to use the Durham test variable.  In
the `seagull diagram' (Fig.~\ref{fig_seagull}) there is a significant
fraction of phase space where the JADE test variable for the two
gluons is less than $\ycut$, whereas those for either gluon and its
associated quark are both greater than $\ycut$.  In this region it
will still be the two gluons that are combined even though this is the
last of the three possibilities to be considered.  As a consequence,
the angular-ordered JADE algorithm still
leads to `phantom jets' in regions where there are no particles,
enhancing the jet multiplicity and hadronization effects.

It is interesting to see that the angular-ordered JADE algorithm with
soft freezing performs even worse in this context than its equivalent without
freezing.  This is because the freezing step increases the mean number
of jets at the hadron level whilst having little effect at the parton level.

\FIGURE[!p]{
\epsfig{file=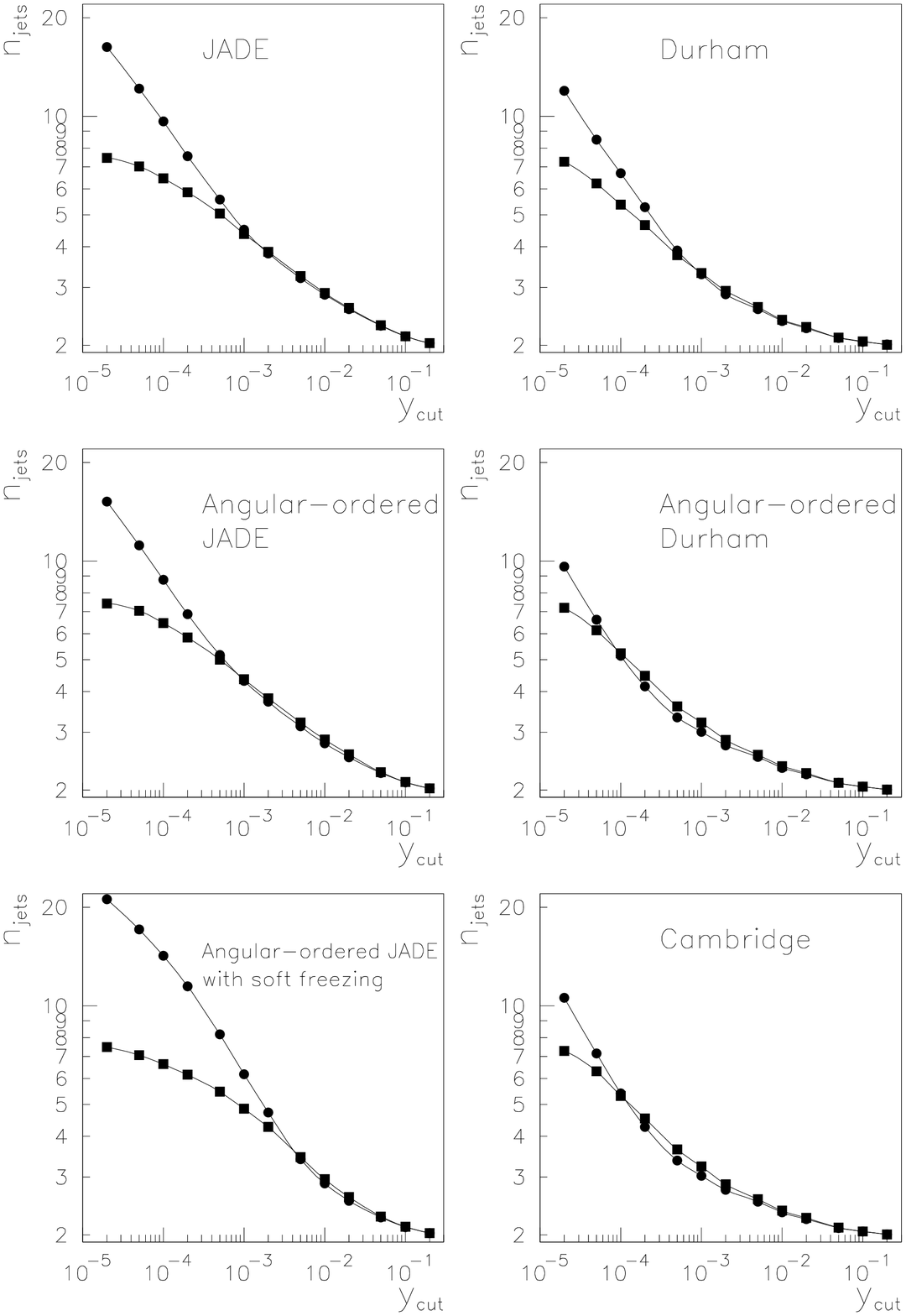,height=20cm}
\caption{Parton- and hadron-level HERWIG results on the mean number of
jets, as in Fig.~\ref{fig_njets90} but at $Q=172$ GeV.}
\label{fig_njets172}
}
Fig.~\ref{fig_njets172} shows the equivalent results at a typical LEP2
energy of $172$ GeV.  It is useful to reduce the hadronization
corrections at this energy since that may help in discriminating
between the process $\ee\to WW \to 4$ jets, used to measure the
$W$ mass, and the background from $\ee\to Z^0/\gamma \to 4$ jets.
Here the results have been continued down to $\ycut = 2\times 10^{-5}$,
the point at which non-perturbative corrections might be expected
to set in at this energy.

The results at $172$ GeV are similar to those at $90$ GeV,
except that they exhibit a lower multiplicity for all six algorithms
at a given $\ycut$, owing to the running of $\as$,
and that the parton- and hadron-level curves diverge at a
smaller value of $\ycut$, owing to the power-decrease in
the hadronization corrections.

\section{Discussion and conclusions}
\label{sec:conc}
In this paper we have identified some deficiencies of the Durham (D)
jet clustering algorithm and their causes. They are due mainly to the
tendency of the algorithm to assemble `junk-jets' from soft, wide-angle
hadrons or partons, and to mis-cluster particles into already-resolved
soft, wide-angle jets. To suppress junk-jet formation, we propose
the separation of the ordering and resolution testing steps of the algorithm.
Using the relative angle as the ordering variable and the $\kt$-resolution
as the test variable, we obtain the angular-ordered Durham (A) algorithm.
To reduce mis-clustering, we propose `soft freezing', i.e.\ forbidding
further clustering with the softer jet when a pair of jets
are resolvable but have the smallest angle. Together with
angular ordering, this gives the Cambridge (C) algorithm.

As measures of performance of jet algorithms, we studied two
quantities in a simple `tube' model of hadronization:
$\VEV{y_3}$, the mean value of the resolution at which
three jets are just resolved, and $\VEV{n_3}$, the mean
multiplicity in the just-resolved third jet. Since the tube
model has no real multijet production, in an optimal algorithm
these quantities should both be as small as possible.  We found
that $\VEV{y_3}$ decreases much more rapidly with increasing
energy in the Durham algorithm than in the original JADE scheme.
Nevertheless the Durham value is enhanced by junk-jet formation, and
$\VEV{y_3}$ is therefore reduced significantly by the angular-ordering
step in the A and C schemes. The quantity $\VEV{n_3}$,
on the other hand, is quite similar in the JADE and Durham schemes,
smaller but still increasing in the A scheme, and essentially
flat in the C scheme. Thus the two measures provide neatly
complementary evidence of the benefits of the two new features
of the C algorithm.

Next we presented full NLO perturbative calculations of the jet fractions
in the new A and C algorithms, and compared the results with those
for the original D scheme. The main difference is that the angular
ordering in the new algorithms shifts a part of the three-jet rate into
the two-jet fraction, while the extra `soft freezing' in the C
scheme moves some of the three-jet rate into the four-jet one.
This is understandable because angular ordering tends to permit
more clustering, while soft freezing prevents some clustering.
The beneficial results include a significant reduction in the
sensitivity to the renormalization scale.

We also extended the parametrization of the fixed-order results for
the D scheme in Ref.~\cite{BKSS} down to $\ycut=10^{-3}$ and provided
similar parametrizations for the A and C schemes.

For a more realistic estimate of hadronization effects than that
obtained from the tube model, we investigated the mean number of
jets as a function of jet resolution in the region of low $\ycut$.
We combined the fixed-order results for this quantity
with those obtained by resummation of leading and
next-to-leading logarithms of $\ycut$ to all orders,
thereby obtaining reliable perturbative predictions
for all $\ycut<10^{-1}$.  The resummed
terms are the same for the D, A and C algorithms;  only the
subleading logarithms and non-logarithmic parts had to be fitted
differently in each case. This extended the results of Ref.~\cite{CDFW1}
for the jet multiplicity in the D scheme to include the A and C schemes.

Finally, we used the Monte Carlo program HERWIG to generate simulated
$\ee$ annihilation events, and compared the parton- and hadron-level
jet multiplicities for the Durham algorithm and its proposed modifications
with the perturbative predictions and with each other.
We also investigated the equivalent modifications to the JADE algorithm,
but these either showed no benefit or performed worse than the original.

We found that the mean number of jets at the HERWIG parton level
was similar in the D, A and C schemes, and close to that predicted
perturbatively, albeit for a rather high effective value of $\as$.
The HERWIG hadron-level values were more different, being lowest for
the A scheme, slightly higher for the C scheme, and much higher for
the original D scheme at small $\ycut$. Down to $\ycut\sim 10^{-3}$
at $\Ecm=M_Z$ ($10^{-4}$  at $\Ecm=172$ GeV), the C-scheme hadron-level
values lie close to the perturbative ones, if the HERWIG input parameter
{\tt QCDLAM} is interpreted as the QCD scale parameter $\lms$. In addition,
the hadronization effects are greatly reduced below $\ycut\sim 10^{-3}$,
suggesting that the C algorithm will be particularly useful for exploring
the interface between perturbative and non-perturbative
dynamics \cite{DLMWprep}. 

A Fortran subroutine to perform jet clustering according to the C
algorithm, {\tt CAMJET}, may be obtained via the World-Wide Web at\\
\href{http://www.hep.phy.cam.ac.uk/theory/webber/camjet/camjet.html}
{{\tt http://www.hep.phy.cam.ac.uk/theory/webber/camjet/camjet.html}}.

\acknowledgments

We have benefited from valuable conversations on this
topic with S.\ Bethke, S.\ Catani and M.H.\ Seymour.
We are grateful to Nigel Glover for his help in using the program EERAD. 
Yu.L.D.\ thanks the Cavendish Laboratory and B.R.W.\ thanks
the CERN Theory Division and the St Petersburg
Nuclear Physics Institute for hospitality
while part of this work was carried out.

This research was
supported in part by the U.K.\ Particle Physics and
Astronomy Research Council and by the EC Programme
``Training and Mobility of Researchers", Network
``Hadronic Physics with High Energy Electromagnetic Probes",
contract ERB FMRX-CT96-0008.

\goodbreak


\begin{thebibliography}{99}
\bibitem{JADE}
       JADE Collaboration, W.\ Bartel et al., \pl{123}{460}{83};
       \zp{33}{23}{86}.
\bibitem{Durham}
       Yu.L.\ Dokshitzer, contribution cited in Report of the Hard QCD
       Working Group, Proc.\ Workshop on Jet Studies at LEP and HERA,
       Durham, December 1990, \jpg{17}{1537}{91}.
\bibitem{BS1} 
%JET CROSS-SECTIONS AT LEADING DOUBLE LOGARITHM IN E+ E- ANNIHILATION
       N.\ Brown and W.J.\ Stirling, \pl{252}{657}{90}.
\bibitem{Stef}
%JET TOPOLOGY AND NEW JET COUNTING ALGORITHMS
       S.\ Catani, in ``QCD at 200-TeV", Proc.\ 17th Workshop of the
       INFN Eloisatron Project, Erice, June 1991.
\bibitem{CDOTW}
%NEW CLUSTERING ALGORITHM FOR MULTI - JET CROSS-SECTIONS IN E+ E- ANNIHILATION
       S.\ Catani, Yu.L.\ Dokshitzer, M.\ Olsson, G.\ Turnock and B.R.\
       Webber, \pl{269}{432}{91}. 
\bibitem{BS2}
%FINDING JETS AND SUMMING SOFT GLUONS: A NEW ALGORITHM
       N.\ Brown and W.J.\ Stirling, \zp{53}{629}{92}. 
\bibitem{Jetset}
       T.\ Sj\"ostrand, \cpc{28}{229}{83}, \ibid{39}{84}{347};\\
       M.\ Bengtsson and T.\ Sj\"ostrand, \ibid{43}{87}{367}.
\bibitem{Herwig}
  G.\ Marchesini, B.R.\ Webber, G.\ Abbiendi, I.G.\ Knowles,
  M.H.\ Seymour and L.\ Stanco, \cpc{67}{465}{92}.
\bibitem{Opalclus}
        OPAL Collaboration, M.Z.\ Akrawy et al., \zp{49}{375}{91}.
\bibitem{BKSS}
%NEW JET CLUSTER ALGORITHMS: NEXT-TO-LEADING ORDER QCD AND
%HADRONIZATION CORRECTIONS
        S.\ Bethke, Z.\ Kunszt, D.E.\ Soper and W.J.\ Stirling,
        \np{370}{310}{92}. 
\bibitem{DKT96}
        Yu.L.\ Dokshitzer, V.A.\ Khoze and S.I.\ Troyan, \pr{53}{89}{96}.
\bibitem{DLMWprep}
        Yu.L.\ Dokshitzer, G.D.\ Leder, S.\ Moretti and B.R.\ Webber,
        Cambridge preprint in preparation.
\bibitem{CDFW1}
%{\em The Average Number of Jets in $e^+e^-$ Annihilation},
     S.\ Catani, Yu.L.\ Dokshitzer, F.\ Fiorani and B.R.\ Webber,
     \np{377}{445}{92}.
\bibitem{OPALnj}
OPAL Collaboration, P.D.\ Acton et al., \zp{59}{1}{93}.
\bibitem{BPY}
       Yu.L.\ Dokshitzer, G.\ Marchesini and B.R.\ Webber,
       \np{469}{93}{96}.
\bibitem{AO}
 A.H.\ Mueller, \pl{104}{161}{81};\\
 B.I.\ Ermolaev and V.S.\ Fadin, \jetpl{33}{81}{285}.
\bibitem{KTCLUS}
%SEARCHES FOR NEW PARTICLES USING CONE AND CLUSTER JET
%ALGORITHMS: A COMPARATIVE STUDY
 M.H.\ Seymour, \zp{62}{127}{94} and private communications;\\
{\tt KTCLUS} program available at
\href{http://surya11.cern.ch/users/seymour/ktclus/}
{{\tt http://surya11.cern.ch/users/seymour/ktclus/}}.
\bibitem{y3resum}
 S. Catani, G. Turnock, B.R. Webber and L. Trentadue, 
 preprint Cavendish-HEP-90/16, August 1990 (unpublished);
 \np{407}{3}{93};\\
 G.\ Dissertori and M. Schmelling, \pl{361}{167}{95}.
\bibitem{numa3} A. Signer, preprint SLAC-PUB-7531, submitted to
{\it Comp.\ Phys.\ Commun}. 
\bibitem{slac} Z. Bern, L. Dixon, D.A. Kosower and S. Weinzier,
\np{489}{3}{97};\\
Z. Bern, L. Dixon, D.A. Kosower, preprint SLAC-PUB-7529, June 1996,
\hepph{9606378};\\
A. Signer and  L. Dixon, \prl{78}{811}{97};\\
A. Signer, SLAC preprint SLAC-PUB-7490, May 1997, \hepph{9705218};\\
L. Dixon and A. Signer, 
preprint SLAC-PUB-7528, June 1997, \hepph{9706285}.
\bibitem{durham} E.W.N. Glover and D.J. Miller, 
\pl{396}{257}{97};\\
J.M. Campbell, E.W.N. Glover and D.J. Miller,
preprint DTP/97/44, RAL TR 97-027, June 1997, \hepph{9706297}.
\bibitem{EERAD} W.T. Giele and E.W.N. Glover, \pr{46}{1980}{92}.
\bibitem{BN} F. Bloch and A. Nordsieck, 
{\it Phys.\ Rev.\ }{\bf 52} (1937) 54.
\bibitem{KLN} T. Kinoshita, {\it J. Math. Phys.\ }{\bf 3} (1962) 650;\\
T.D. Lee and M. Nauenberg, {\it Phys.\ Rev.\ }{\bf 133} (1964) 1549.
\bibitem{Sterman} G. Sterman, \pr{17}{2789}{78}. 
\bibitem{5j} A. Ballestrero and E. Maina, \pl{323}{53}{94}.
\bibitem{moriond} A. Ballestrero, E. Maina and S. Moretti,
      Proceedings of the "XXIXth Rencontres de Moriond:
      QCD and High Energy Hadronic Interactions", M\'eribel, Savoie, France,
      16/26 March 1994, ed. by J. Tr\^an Thanh V\^an, ed.
      Fronti\`eres, Gif-sur-Yvette, 1994, 367.
\bibitem{as} The LEP Electroweak Working Group and the SLD Heavy Flavor
             Group, {preprints} CERN-PPE-96-183, December 1996.
\bibitem{subtraction} G. Grunberg, \pl{95}{70}{80};\\
S.J. Brodsky, G.P. Lepage and P.B. Mackenzie, \pr{28}{228}{83};\\
P.M. Stevenson, \np{231}{65}{84};\\
H.D. Politzer, \np{194}{493}{82}.
%\bibitem{thrust} S. Brandt, Ch. Peyrou, R. Sosnowski and A. Wroblewski,
%{Phys. Lett.}  (1964) 57;\\
%E. Farhi, \prl{39}{1587}{77}.
\bibitem{CDW2}
%{\em Average Number of Jets in Deep Inelastic Scattering}
     S.\ Catani, Yu.L.\ Dokshitzer and B.R.\ Webber, \pl{322}{263}{94}.
\bibitem{HW59}
     G.\ Marchesini, B.R.\ Webber, G.\ Abbiendi, I.G.\ Knowles,
     M.H.\ Seymour and L.\ Stanco, \hepph{9607393}.
\bibitem{CMW}
     S.\ Catani, G.\ Marchesini and B.R.\ Webber, \np{349}{635}{91}.
\end{thebibliography}
\end{document}